%% file: main.tex
\documentclass[fleqn,usenatbib]{mnras}

\usepackage{mathptmx}

\usepackage[T1]{fontenc}
\usepackage{ae,aecompl}

\usepackage{rotating}
\usepackage{booktabs}
\usepackage{setspace}
\usepackage{txfonts}
\usepackage{float}
\usepackage{amssymb}
\usepackage{epsfig}
\usepackage{color}
\usepackage{graphicx}
\usepackage{multirow}
\usepackage{longtable}
\usepackage{verbatim}
\usepackage[flushleft]{threeparttable}
\usepackage{gensymb}
\usepackage{epstopdf}
\usepackage{array}
\usepackage{natbib}
\usepackage{xspace}
\usepackage{enumitem}

\graphicspath{{./}{figures/}}
\newcommand{\Ni}{\emph{NICER}\xspace}
\newcommand{\sw}{\emph{Swift}\xspace}
\newcommand{\wise}{\emph{WISE}\xspace}
\newcommand{\neowise}{\emph{NEOWISE}\xspace}
\newcommand{\avd}{AT~2019avd\xspace}

\title[The radio detection and accretion properties of AT~2019avd]{\centering The radio detection and accretion properties of the peculiar nuclear transient {\avd}}

\author[Y. Wang et al.]{\centering
Yanan Wang$^{1}$\thanks{E-mail: Y.Wang@soton.ac.uk},
Ranieri D. Baldi$^{2,1}$,
Santiago del Palacio$^{3}$,
Muryel Guolo$^{4}$,
Xiaolong Yang$^{5,14}$,\newauthor
Yangkang Zhang$^{5}$,
Chris Done$^{6}$,
Noel Castro Segura$^{1}$,
Dheeraj R. Pasham$^{7}$,
Matthew Middleton$^{2}$,\newauthor
Diego Altamirano$^{1}$,
Poshak Gandhi$^{1}$,
Erlin Qiao$^{8}$,
Ning Jiang$^{9}$,
Hongliang Yan$^{8}$,
Marcello Giroletti$^{2}$,\newauthor
Giulia Migliori$^{2}$,
Ian McHardy$^{1}$,
Francesca Panessa$^{10}$,
Chichuan Jin$^{8,11}$,
Rongfeng Shen$^{12}$
and Lixin Dai$^{13}$
\\
$^{1}$Physics \& Astronomy, University of Southampton, Southampton, Hampshire SO17~1BJ, UK\\
$^{2}$INAF - Istituto di Radioastronomia, Via P. Gobetti 101, I-40129 Bologna, Italy\\
$^{3}$Department of Space, Earth and Environment, Chalmers University of Technology, SE-412 96 Gothenburg, Sweden\\
$^{4}$Department of Physics and Astronomy, Johns Hopkins University, 3400 N. Charles St., Baltimore MD 21218, USA\\
$^{5}$Shanghai Astronomical Observatory, Key Laboratory of Radio Astronomy, Chinese Academy of Sciences, Shanghai 200030, China\\
$^{6}$Centre for Extragalactic Astronomy, Department of Physics, University of Durham, South Road, Durham DH1 3LE, UK\\
$^{7}$Kavli Institute for Astrophysics and Space Research, Massachusetts Institute of Technology, Cambridge MA 02139, USA\\
$^{8}$National Astronomical Observatories, Chinese Academy of Sciences, 20A Datun Road, Beijing 100101, China\\
$^{9}$CAS Key laboratory for Research in Galaxies and Cosmology, Department of Astronomy, University of Science and Technology of China, Hefei 230026, China\\
$^{10}$INAF - Istituto di Astrofisica e Planetologia Spaziali, via del Fosso del Cavaliere 100, I-00133 Roma, Italy\\
$^{11}$School of Astronomy and Space Sciences, University of Chinese Academy of Sciences, 19A Yuquan Road, Beijing 100049, China\\
$^{12}$School of Physics and Astronomy, Sun Yat-Sen University, Zhuhai 519082, China\\
$^{13}$Department of Physics, University of Hong Kong, Pokfulam Road, Hong Kong, China\\
$^{14}$Shanghai Key Laboratory of Space Navigation and Positioning Techniques, Shanghai Astronomical Observatory, CAS, Shanghai 200030, China\\
}

\date{Accepted XXX. Received YYY; in original form ZZZ}

\pubyear{2022}

\begin{document}
\label{firstpage}
\pagerange{\pageref{firstpage}--\pageref{lastpage}}
\maketitle

\begin{abstract}
AT~2019avd is a nuclear transient detected from infrared to soft X-rays, though its nature is yet unclear. The source has shown two consecutive flaring episodes in the optical and the infrared bands and its second flare was covered by X-ray monitoring programs. During this flare, the UVOT/\sw photometries revealed two plateaus: one observed after the peak and the other one appeared $\sim240$\,days later. Meanwhile, our \Ni and XRT/\sw campaigns show two declines in the X-ray emission, one during the first optical plateau and one 70--90\,days after the optical/UV decline. The evidence suggests that the optical/UV could not have been primarily originated from X-ray reprocessing. 
Furthermore, we detected a timelag of $\sim$16--34\,days between the optical and UV emission, which indicates the optical likely comes from UV reprocessing by a gas at a distance of $0.01-$0.03\,pc.
We also report the first VLA and VLBA detection of this source at different frequencies and different stages of the second flare. The information obtained in the radio band -- namely a steep and a late-time inverted radio spectrum, a high brightness temperature and a radio-loud state at late times -- together with the multiwavelength properties of \avd suggests the launching and evolution of outflows such as disc winds or jets.
In conclusion, we propose that after the ignition of black hole activity in the first flare, a super-Eddington flaring accretion disc formed and settled to a sub-Eddington state by the end of the second flare, associated with a compact radio outflow.

\end{abstract}




\section{Introduction}
The sample of nuclear transients has shot up in recent years as more survey telescopes have been put into operation, such as the Zwicky Transient Facility (ZTF; \citealt{Bellm2019}) and the all sky automated survey super nova (ASASSN; \citealt{Shappee2014}) in optical bands, and \textit{eROSITA} \citep{Predehl2021} in soft X-rays. Active galactic nuclei (AGN) generally show low variability on long timescales spanning from months to years. Transients discovered in the vicinity of supermassive black holes (SMBHs) instead show more extreme phenomenology, e.g. `changing-look' AGN (CLAGN; e.g. \citealt{Matt2003,LaMassa2015,Ricci2021,Guolo2021}), Narrow-line Seyfert 1 galaxies \citep[NLSy1; e.g.][]{boller96,komossa07}, tidal disruption events (TDEs; e.g. \citealt{Komossa1999,Velzen2021,Hammerstein2022}) and supernovae (SNe; e.g. \citealt{Ulvestad1997,Mattila2001,Villarroel2017}). 
Additionally, some other types of nuclear transients that show exotic phenomenology have also been discovered, such as quasi-periodic eruptions with high-amplitude X-ray bursts recurring every few hours/days \citep{Miniutti2019,Giustini2020,Arcodia2021}, and
sudden ignition of optical-UV emission with strong double-peaked emission lines and with a decline rate much slower than in TDEs \citep{Tadhunter2017,Trakhtenbrot2019,Gromadzki2019}.

A TDE has been predicted to occur when a star passes inside the tidal radius of a SMBH \citep{Hills1975,Rees1988}. Such events can result in a months-to-years-long flare with an extreme multiwavelength variability (see \citealt{Gezari2021} for a recent review on TDEs). Although the first few TDEs were detected by {\it ROSAT} in X-rays \citep{Komossa1999}, most of the TDEs discovered in the literature are UV/optically selected. 
In only around a dozen cases, they are bright in both optical and X-ray bands, with an X-ray-to-optical luminosity ratio around the peak of the X-ray emission of either $\sim1$ or $>1$ when a jet is present \citep{Auchettl2017}.
In other cases, they are bright either in optical/UV bands (e.g. PS1--10jh, \citealt{Gezari2012}; ASAS-SN14ae, \citealt{Holoien2014}) or in X-rays exclusively, or they peak in optical/UV bands with faint/no X-ray detection first but later peak/brighten in X-rays with a time delay of months to years (e.g. D3--13, \citealt{Gezari2006}; ASASSN-14li, \citealt{2017ApJ...837L..30P}; ASASSN--15oi, \citealt{Gezari2017}; OGLE16aaa, \citealt{Shu2020}; AT2019azh/ASASSN-19dj, \citealt{Hinkle2021}).
The latter case results in a large optical-to-X-ray ratio of 1 to 1000 -- or even higher when no X-ray emission is detected -- that significantly evolves with time \citep{Gezari2017}.

Such diversity of TDE flaring phenomena have motivated several theoretical models.
\cite{Metzger2016} proposed a reprocessing model in which the X-ray emission from the inner disc is mostly suppressed at early times and the X-rays can only escape after the obscuring material is completely ionised. 
Alternatively, \cite{Gezari2017} suggested that the late-time brightening of the X-ray emission could be due to the delayed accretion through a newly forming debris disc.
\cite{Dai2018} and \cite{Curd2019} emphasised the viewing angle effects on TDE flares, suggesting X-ray bright TDEs are observed through a low-density funnel region whereas the optical/UV bright TDEs are observed through the disc edge where X-rays are reprocessed into optical/UV emission by the outer disc or the optically thick outflows.
Furthermore, \cite{Wen2020} proposed a slimming disc with a near edge-on configuration whose X-ray emission increases as more of the inner disc region is exposed to the observer.
Recently, \cite{Mummery2021} proposed a unified model of disc-dominated TDEs and they showed that the peak Eddington ratio of the disc is a decisive parameter of the observed properties of TDEs.

Radio and infrared (IR) emission have also been detected in follow-up observations of TDEs (or candidates). For the former, these sources show a wide range of radio luminosities of $10^{37}-10^{42}\,\rm erg\,s^{-1}$, which can originate from jets (\citealt{Zauderer2011,Bloom2011,Berger2012,Cenko2012,Irwin2015,Mattila2018,Pasham2018}), winds (\citealt{Alexander2016}) or unbound tidal debris streams interacting with the surrounding circumnuclear medium (\citealt{Krolik2016,matsumoto21}; also see \citealt{Alexander2020} for a review). 
For the latter, IR emission is generally interpreted as optical/UV/soft X-ray radiation being reprocessed by dust within a few sub-pc of the central SMBH, with a covering factor of $\sim0.01$ \citep{Velzen2016,Jiang2016,Jiang2021,Velzen2021}. The dust echo are an important tool for studying the circumnuclear environment of galaxies, down to sub-pc scales, and for measuring the bolometric luminosity ($L_\mathrm{bol}$) of TDEs \citep[][]{Velzen2016,Gezari2021}.
 
The diversity of multi-band emitting TDEs has also complicated their distinction from other sources such as SNe and active SMBHs \citep{zabludoff21}. 
SNe can occur in the centre of a galaxy as well \citep{Ulvestad1997,Mattila2001,Villarroel2017}, and variable AGNs -- such as CLAGN and NLSy1 -- may also show explosive changes in luminosity on timescales of months to years \citep{Yan2019,frederick21}. 
The identification of TDEs may be even more difficult when a TDE takes place in a pre-existing AGN (see \citealt{Chan2019} for hydrodynamic simulations of a debris stream colliding with a pre-existing accretion disc; \citealt{Ricci2020}).

\subsection{The transient: {\avd}}

{\avd} (also known as eRASStJ082337+042303 or ZTF19aaiqmgl), located at $z=0.028$, is a nuclear transient that flared in IR, optical/UV and X-ray wavelengths. Its optical emission was first detected by ZTF on February 9, 2019 \citep{Nordin2019}, spatially associated with the quiescent galaxy 2MASX~J08233674+0423027 \citep{Alam2015}. The long-term ZTF lightcurve shows two extraordinary consecutive optical flaring episodes lasting over two years in total. The X-ray flare was first detected by SRG/{\it eROSITA} on April 28, 2020 during the rising phase of the second optical flare \citep{Malyali2020}. 
The concurrent optical and X-ray luminosity evolution (with multiple rising and drops) of the source makes it a rather unique nuclear transient. The source shows evolving optical spectral lines and Bowen fluorescent lines \citep{Malyali2021}. The SMBH mass estimated from empirical relations (virial mass method and host-BH mass relation) and from the single-epoch mass-estimation technique are 10$^{6.1-7.2}\,M_{\odot}$ \citep{frederick21} and $10^{6.3\pm0.3}\,M_{\odot}$ \citep{Malyali2021}, respectively. We will adopt the latter value throughout this work.

Part of the IR, optical/UV and X-ray data from Wide-field Infrared Survey Explorer \citep[{\wise},][]{Wright_2010}, ZTF, UVOT/{\sw} and XRT/{\sw} campaigns of \avd have been reported in \cite{Malyali2021,frederick21} and \cite{Chen2022}. \cite{Malyali2021} concluded that its X-ray properties are consistent with TDEs but the optical/UV emission are not TDE-like. \cite{frederick21} argued against a TDE origin based on its optical photometry and, instead, suggested a flaring NLSy1 type of transient with significant He~{\sc ii} and N~{\sc iii} profiles. \cite{Chen2022} proposed a two phase model in which {\avd} is caused by the partial disruption of a star by a SMBH, and the two optical flares are due to stream circularization and delayed accretion, respectively.
Here we include new and archival observations from ZTF, {\sw} and {\wise}, together with new data from \Ni
\footnote{We only include the \Ni lightcurve in this work. A detailed study of the temporal and spectral properties of the X-rays of \avd with the \Ni data will be presented in Wang et al. (in prep).}.
More importantly, we also report the first radio detection of this target with the Karl Guthe Jansky Very Large Array (VLA) and Very Long Baseline Array (VLBA).

\begin{figure}
\centering  
\includegraphics[width=0.9\linewidth]{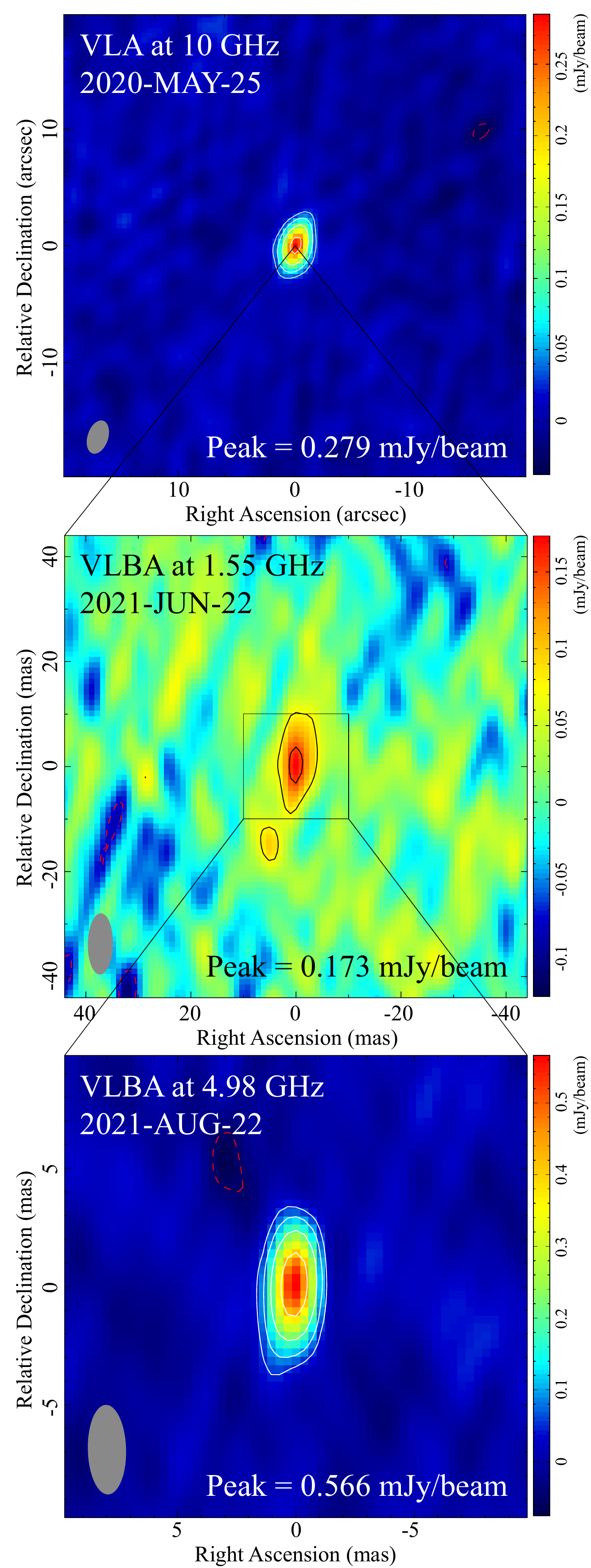}
\vspace{-0.1cm}
\caption{High-resolution images of {\avd} derived from the VLA X band (top panel) and VLBA L and C bands (middle and bottom panels, respectively) observations. The image information (beam sizes, rms, etc.) is given in Table~\ref{radiotab}. The positive contours are plotted in white and black, and the negative contours are plotted in red. Contour levels are $3\sigma\times(-1, ~1, ~2, ~4,~8)$, with $\sigma$ the rms noise.}
\label{fig:vlbaimg}
\end{figure}

\section{Multiwavelength observations and data reduction}
We adopt a flat $\Lambda$CDM cosmology with $H_{0}=67.4\,\rm km\,s^{-1}Mpc^{-1}$ and $\Omega_{\rm m}=0.315$ from \cite{Planck2020}, which implies a luminosity distance of $D \sim 130$\,Mpc.

\subsection{ZTF}
ZTF monitored \avd in the {\it r} and the {\it g} bands from January 12, 2019 to May 4, 2021. The ZTF lightcurves are downloaded from the Lasair alert  broker\footnote{\url{https://lasair.roe.ac.uk/object/ZTF19aaiqmgl/}}. We apply Galactic extinction correction on both bands using $E(B-V)= 0.022$ \citep{Schlafly2011}.

\subsection{UVOT}
{\sw} performed 51 target-of-opportunity (ToO) observations on this target from May 13, 2020 to May 26, 2022, which add up to a total exposure time of 56.4\,ks with good time intervals. 
The task {\sc uvotimsum} was applied to sum all the exposures when more than one snapshot was included in each individual filter data. 
The task {\sc uvotsource} was run to extract magnitudes from aperture photometry. To allow a consistent subtraction of the host contribution to the UVOT bands, a circular region of the size of the host's Petrosian radius, 10\arcsec, centred at the target position was chosen for the source and another region of 40\arcsec~located at a nearby position was used to estimate the background emission. The target was observed with all/some of the six filters, UVW2 (central wavelength, 1928\,\AA), UVM2 (2246\,\AA), UVW1 (2600\,\AA), U (3465\,\AA), B (4392\,\AA) and V (5468\,\AA) in different observations. The UVOT magnitudes have been host-subtracted (see Appendix~\ref{sec:app}), and corrected for Galactic extinction.

\subsection{\wise}

To investigate the mid-infrared (MIR) variability of \avd, we employed photometric data of the W1 ($3.4~\mu$m) and W2 ($4.6~\mu$m) bands from {\wise}. Multi-epoch data was gathered from the reactivation \neowise mission \citep{Mainzer_2014} database using the \neowise-R Single Exposure (L1b) Source Table on the NASA/IPAC Infrared Science Archive (IRSA\footnote{\url{https://irsa.ipac.caltech.edu/}}) with a matching radius of 2\arcsec. \neowise visits a particular field in the sky every six months and at each visit three observations are performed, each a couple of days apart. We binned the photometric data into $\sim\pm$15\,days to obtain a representative magnitude in each visit. To remove poor-quality data, only the photometry flagged with $cc\_flags=0$, $qual\_frame>0$, $qi\_fact>0$, $saa\_sep>0$, and $moon\_masked=0$ were averaged after 3$\sigma$ clipping in a single visit. Following \cite{Lyu_2019}, we evaluated the magnitude uncertainty for each epoch as follows:
\begin{equation}
\sigma^2_\mathrm{epoch} = \frac{1}{N-1} \sum_{i=1}^{N} \left(m_i - \overline{m_\mathrm{epoch}}\right)^2 + \frac{1}{N^2}\sum_{i=1}^{N}\sigma^2_{i,\mathrm{pho}} + \frac{1}{N} \sigma^{2}_\mathrm{ss},
\end{equation}
where $m_i$ and $\sigma_{i,\mathrm{pho}}$ denote the magnitude and its uncertainty at each observation, respectively, in the Vega magnitude system; $\overline{m_\mathrm{epoch}}$ denotes the mean magnitude at each epoch; and $\sigma_\mathrm{ss}$ denotes the system stability ($\sim$ 0.016 mag for \neowise). 

\input{table_radio}

\subsection{VLA}
The transient was observed in the X-band with the VLA in C-array configuration on 25th May 2020 (Project 20A-514; PI: Baldi), during the second optical flaring episode. The target was observed at the central frequency of 10\,GHz and with a bandwidth of 4\,GHz for 18\,min bracketed between scans of the secondary (phase) calibrator, which was observed for 1.5\,min. The scans of the absolute flux density scale calibrator (3C~138) were performed at the end of the scheduling block for 4.5\,min. The data calibration and reduction procedure were performed with the calibration pipeline within the Common Astronomy Software Application (\textsc{casa 5.4.1 version}, \citealt{mcmullin07}). After calibration, the plots were inspected for residual interference. For the image reconstruction, the {\sc tclean} task in \textsc{casa} was used. 
We used the \texttt{mtmfs} deconvolver mode, which allows to reconstruct images from visibilities using a multi-term (multi-scale) multi-frequency approach \citep{rau11} with 2 Taylor coefficients in the spectral model (\texttt{nterms=2}). 
We produced full resolution maps considering the \cite{Briggs1995} initial weighting algorithm with robustness parameter equal to 0.5, which ensures a balance between resolution and sensitivity. The restoring clean beam size is 3.2$\arcsec \times $1.7\arcsec\, and the final radio map reaches an rms of 8.5\,$\mu$Jy\,beam$^{-1}$. 
The top panel of Fig.~\ref{fig:vlbaimg} shows the 10-GHz map of {\avd} and reveals an unresolved component with a peak brightness $279.2\pm5.5\,\mu$Jy\,beam$^{-1}$. Further details are provided in Table~\ref{radiotab}.

Due to the low statistics it was not possible to extract an in-band spectral index image using \textsc{casa}. Therefore we split the 4-GHz bandwidth of the X-band dataset into 2 sets of visibilities 2-GHz wide centred at 9.04 and 11.06\,GHz. 
We also tried to further divide the 4-GHz bandwidth in 3 frequency bins centred at 8.6, 9.9 and 11.3\,GHz. Unfortunately, large flux uncertainties prevented from deriving an accurate spectrum profile modelled with 
the \textsc{casa} task \textsc{uvmodelfit}, so we do not consider this further splitting reliable and focus only on the former one (9.04-11.06 \,GHz)  .


\subsection{VLBA}

The target was observed by VLBA at L- (1.4\,GHz) and C-band (5\,GHz) (Project BW142A \& B; PI: Wang) during the optical and the X-ray luminosity drops.
The full VLBA antennas includes BR (Brewster), FD (Fort Davis), HN (Hancock), KP (Kitt Peak), LA (Los Alamos), MK (Mauna Kea), NL (North Liberty), OV (Owens Valley), PT (Pie Town) and SC (Saint Croix).
The L-band observation (BW142A) was made on 22th Jun 2021, lasting 2\,hr, during which 9 VLBA antennas participated (OV out due to maintenance). For the C-band observation (BW142B) performed on 22th Aug 2021, all the 10 VLBA antennas joined the 4-hr observation (HN out for about one hour due to the extreme weather). Both observations were recorded at 2\,Gbps data rate (32\,MHz, 8 IF per pol) with dual polarisation.
Phase referencing mode was applied for both epochs and the bright calibrator J0825+0309 (1.35$\degree$ from the target) was used as the phase calibrator. The scheduled phase-referencing nodding cycle was set to 7\,min long with 5\,min on the target and 2\,min on the calibrator, which yields total on-source times of 80\,min and 160\,min at L and C band, respectively.

\begin{figure*}
\centering  
\includegraphics[width=0.9\linewidth]{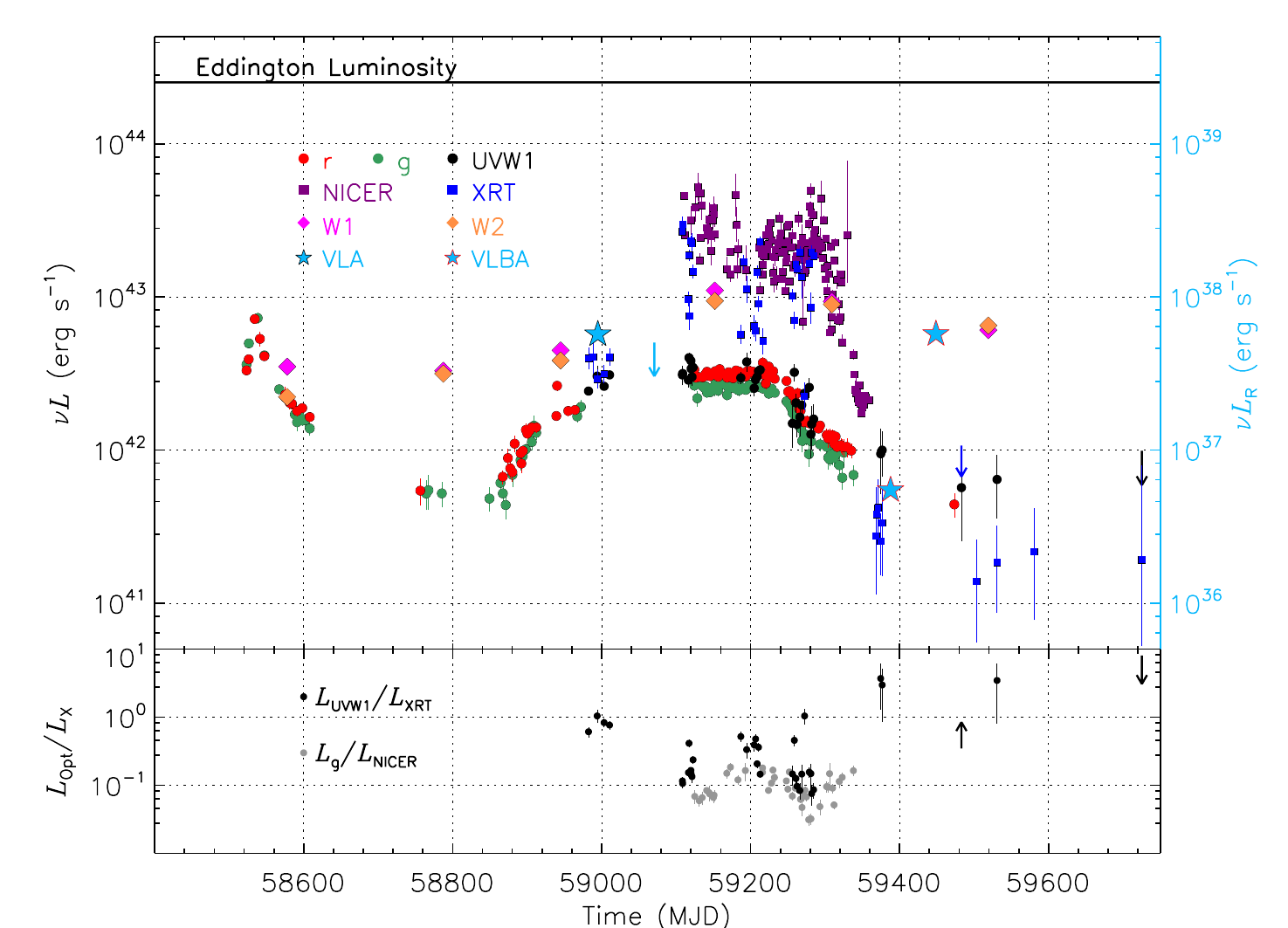}
\caption{\textbf{Upper:} Temporal evolution of multi wavelength luminosity of {\avd}. The left y-axis shows the multi-band luminosities (erg s$^{-1}$) except for the radio-band luminosities which are plotted on the right y-axis. The unabsorbed X-ray luminosities in the 0.3--2\,keV energy range are taken from Wang et al. (in prep). 
The radio data are observed, from left to right, at 10\,GHz and 3\,GHz with VLA, and at 1.6\,GHz and 5\,GHz with VLBA.
\textbf{Lower:} The evolution of the optical-to-X-ray ratio. In both panels the arrows indicate the limit of the detection at 3$\sigma$ confidence level.}
\label{fig:lc}
\end{figure*}

After observation, the data recorded in each antenna were transferred to the DiFX correlators in Socorro, USA \citep{2011PASP..123..275D}, for correlation with an integration time of 2\,s.
After correlation, the visibility data were downloaded by the user and imported into the \textsc{aips} software for further calibration \citep{2003ASSL..285..109G}. The standard calibration procedure of VLBA data were applied according to the \textsc{aips} COOKBOOK\footnote{\url{http://www.aips.nrao.edu/cook.html}}.
At first, the corrections for the ionosphere and Earth Orientation effect were conducted by the \textsc{aips} tasks {\sc vlbatecr} and {\sc vlbaeops}. The \textsc{aips} task {\sc apcal} was performed to calibrate the amplitudes for each antenna taking advantage of the weather condition for the opacity correction. For the phase calibration, a step of manual calibration on the fringe finder scan and a following step of global fringe fitting for the phase calibrator were performed using the \textsc{aips} task {\sc fring}. Finally, the antenna-based bandpass was solved out of the calibrators and applied to the visibility data with the task {\sc bpass}. 

The calibrated visibility data were then exported and imported into the Caltech Difmap software \citep[\textsc{difmap}, ][]{1997ASPC..125...77S} for self-calibration and imaging. We first did \textsc{clean}ing and phase-only self-calibration to image the phase calibrator J0825+0309, generated the final \textsc{clean}ed map and loaded it into \textsc{aips} for another phase calibration {\sc fring}. This step calibrates the residual phase errors caused by the extended structure of the phase calibrator. Then the final phase solutions were interpolated into the target, and \textsc{split}ed for final imaging in \textsc{difmap}. During the imaging of the target source, no self-calibration steps were used due to the weakness of the target source.

The middle and bottom panels of Fig.~\ref{fig:vlbaimg} depict the VLBA maps of {\avd} in L and C bands with angular resolutions of 11.7$\times$4.8~mas$^2$ and 3.7$\times$1.6~mas$^2$, respectively. The source was detected at both 1.6 and 5\,GHz with a signal-to-noise ratio above $5\sigma$ (rms of 25 and 16 $\mu$Jy\,beam$^{-1}$, respectively). The VLBA core position is consistent with the VLA detection (within the VLA beam, see Table~\ref{radiotab}). The VLBA peak brightnesses are $0.173$ and $0.566\,\mathrm{mJy\,beam^{-1}}$ for 1.6 and 5\,GHz, respectively, corresponding to signal-to-noise ratios $\sim7$ and $35$, respectively. The uncertainties in the peak brightness and the total flux density are calculated by combining the systematic uncertainty and the rms value.

\section{multiwavelength temporal evolution} \label{sec:multiwavelength}
The upper panel of Fig.~\ref{fig:lc} shows the long-term evolution of the multiwavelength emission detected with ZTF, {\sw}, {\Ni}, {\wise} and VLA/VLBA. 
ZTF detected two optical flaring episodes in the $r$ and $g$ bands, marked with the red and green dots in the upper panel of Fig.~\ref{fig:lc}: the first one, approximately from MJD~58523 to 58785, shows a sharp peak and the second one, starting from MJD~58849, shows more complex features with a duration at least twice as long as the first one. Specifically, the second flare shows a plateau of nearly constant luminosity for at least 106\,days (from MJD~59124 to 59230). Before the plateau, there is a 5-month observational gap -- from MJD~58972 to 59124, due to Sun constraints -- that makes it unclear when the second flare peaked. After the plateau, the optical luminosity decreases abruptly by a factor of more than three in approximately 106\,days.

The first UVOT observation was triggered about 133\,days (MJD~58982) after the start of the second flare. For clarity, we only show the UVW1 photometry in the upper panel of Fig.~\ref{fig:lc}, which is the most frequently used UV filter in our {\sw} campaign. 
The UVW1 photometry is comparable to the ZTF-r and -g bands, revealing an identical behaviour of the UV to that of the optical emission. However, the B and V magnitudes maintain at the host galaxy level, rather constant during the flare. We show the UVOT photometry in all the six bands in Fig.~\ref{figA:UVOT_ph} with a comparison to the magnitude of the host galaxy. After the seasonal gap in 2021, the late-time UV emission shows a second plateau around MJD~59473 to 59580.

Since MJD~59110, {\Ni} started performing high-cadence monitoring of {\avd} on a nearly daily basis. For both XRT and {\Ni}, the X-ray spectrum is very soft (peaking around 0.1\,keV) and is dominated by the background at energies above 2\,keV. The {\Ni} and XRT luminosities -- shown as purple and blue squares, respectively, in the upper panel of Fig.~\ref{fig:lc} -- are taken from Wang et al. (in prep), who obtained the fluxes by fitting the {\Ni} and the XRT spectra. Due to the same seasonal gap around the peak period of the optical flare, it is unclear whether we have captured the peak of the flare in X-rays. However, the X-ray luminosity increased over one order of magnitude after the gap, reaching $L_{\rm X} \sim 6.5\times10^{43}\,\rm erg\,s^{-1}$ on MJD~59110. 

Later, the X-ray luminosity temporarily decreased by a factor of 6 in 100\,days and increased by a factor of 3 in the next 120\,days. It then decreased significantly by over one order of magnitude since MJD~59300, roughly 70\,days after the drop in the optical flare. 
Before another seasonal gap between MJD~59400 and 59500, {\sw} was able to monitor {\avd} for a few days more than \Ni without the contamination from the Sun glare. Compared to the X-ray peak on MJD~59110, the X-ray luminosity on MJD~59368--59377 dropped by over two orders of magnitude, which is consistent with the luminosity observed after the gap. 

It is clear from Fig.~\ref{fig:lc} that the optical decay starts earlier than the X-ray decline, with the latter exhibiting a faster decay. To quantify the delay between the optical and X-ray decay, we use the discrete correlation function \citep[DCF;][]{EdelsonKrolik:1988} 
and use the bootstrapping technique \citep[e.g. ][]{astroML_boook} to determine the error. 
In this calculation we have only included the \Ni and ZTF data between MJD~59100 and 59338, quoted values for the lag and its confidence interval are the median and the 16th and 84th quantile of the 5000 trials bootstrapped distribution. 
This yields a lag of the X-rays with respect to the optical bands of $\tau_{\rm g-X} = 78^{+9}_{-12}\,\rm{d}$ and $\tau_{\rm r-X} = 67^{+15}_{-18}\,\rm{d}$ for the $g$ and $r$ filters, respectively. 

To study the relative changes of the optical/UV and the X-ray flares, we calculated the optical/UV-to-X-ray ratio with the data of either ZTF and {\Ni} or UVW1 and XRT, respectively. As shown in the bottom panel of Fig.~\ref{fig:lc}, the luminosity ratio varies between 0.03 and 0.19 during the main body of the flare from MJD~59110 to 59338; this ratio is $\sim$1 in the period prior to the peak and increases to up to $\sim$10 at late times. These ratios are consistent with those obtained by measuring the fluxes from fitting the X-ray and optical-UV SEDs.

In the radio band, we searched for pre-flare radio observations from public radio surveys at different frequencies and angular resolutions, namely the Faint Images of the Radio Sky (FIRST, \citealt{becker95}), NRAO VLA Sky Survey (NVSS, \citealt{condon98}), TIFR GMRT Sky Survey (TGSS, \citealt{intema17}), VLA Sky Survey (VLASS, \citealt{lacy20}) and LOFAR Two-metre Sky Survey (LOTSS, \citealt{Shimwell2017}). The source was not detected in any of these radio surveys prior to the 2019 optical flare: the VLA surveys, FIRST in 2001 and VLASS in 2017, give a 3$\sigma$ upper limit of the radio flux density of 0.44\,mJy\,beam$^{-1}$ at 1.4\,GHz and 0.39\,mJy\,beam$^{-1}$ at 3\,GHz, respectively. Then, we detect radio emission in the position of the transient at the second optical flare with our VLA observation. In addition, there is a VLASS observation 75\,d later (MJD~59070), which reveals the presence of a very weak component whose peak brightness (0.475\,mJy\,beam$^{-1}$) is slightly below the 3$\sigma$ upper limit, 0.506\,mJy\,beam$^{-1}$. Therefore, this result cannot be reported as a statistically significant detection\footnote{We performed a statistical analysis by repeating 10 times the measurement of the putative source and of the background level on different areas to conclude that we cannot make any claim of a robust detection. Therefore, any conclusion on the emission of \avd would be an overinterpretation of this dataset.}. At later times, during the post-flare phase, our VLBA observations reveal a compact core with a flux density at 5\,GHz ($\sim$0.57\,mJy\,beam$^{-1}$) $\sim$3.2 times brighter than the preceding 1.6-GHz component ($\sim$0.17\,mJy\,beam$^{-1}$). 
More details on the radio data are presented in Table~\ref{radiotab}.

\begin{figure} 
\centering  
\includegraphics[width=\linewidth]{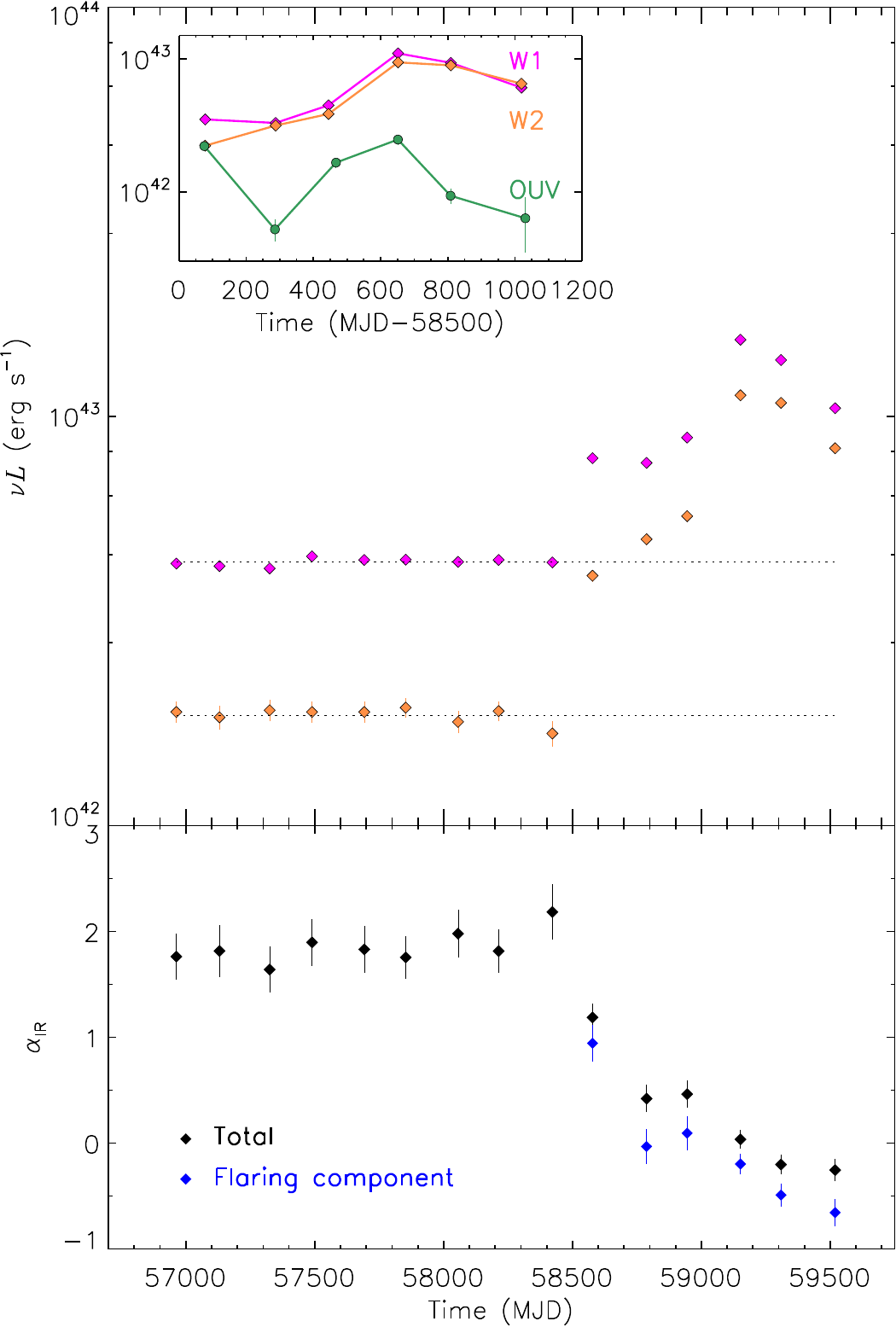}
\vspace{-0.4cm}
\caption{\textbf{Upper:} Temporal evolution of the IR luminosity of {\avd}. 
The lightcurve was separated into a steady component plus a flaring component, the latter most likely associated with {\avd}. 
The dashed line indicates the luminosity level of the steady component prior to the event. The inset shows the quasi-simultaneous evolution of the IR (flaring component) and optical/UV emission.
\textbf{Lower:} The evolution of the spectral index $\alpha_{\rm IR}$ (defined as $F_\nu \propto \nu^\alpha$) determined between W1 and W2 (see Sect.~\ref{sec:multiwavelength}).}
\label{fig:WISE}
\end{figure}

We show the IR lightcurve in the upper panel of Fig.~\ref{fig:WISE}, which spans a baseline of 9\,yr, from December 2013 to March 2022.
The IR flux was very stable from MJD~56700 to 58500, prior to the first optical flare in 2019; after that, the emission suddenly started to increase in both W1 and W2 bands. 
The increase in the IR luminosity by a factor $\sim 2$ was coincident with the onset of the first optical flare. Around MJD~59100 the IR luminosity again increased by an additional factor $\sim 2$, this time coinciding with the second (X-ray) flare from {\avd}. We separate the IR steady component (MJD~56700 to 58500), probably associated with background emission or from the host galaxy, from the flaring component (MJD~58500 to 59520). We quantify the unvarying component by fitting the IR flux before the first optical flare when the source was IR quiet (see the dash line in Fig.~\ref{fig:WISE}, top panel) for W1 and W2. The flaring component is measured by subtracting the constant IR flux to the total emission in the two {\wise} bands. 
We show the relative changes of the IR and the optical/UV luminosities in the inset of Fig.~\ref{fig:WISE} in which we only include the flaring component of the IR and the quasi-simultaneous optical/UV observations.

\begin{figure} 
\centering  
\includegraphics[width=\linewidth]{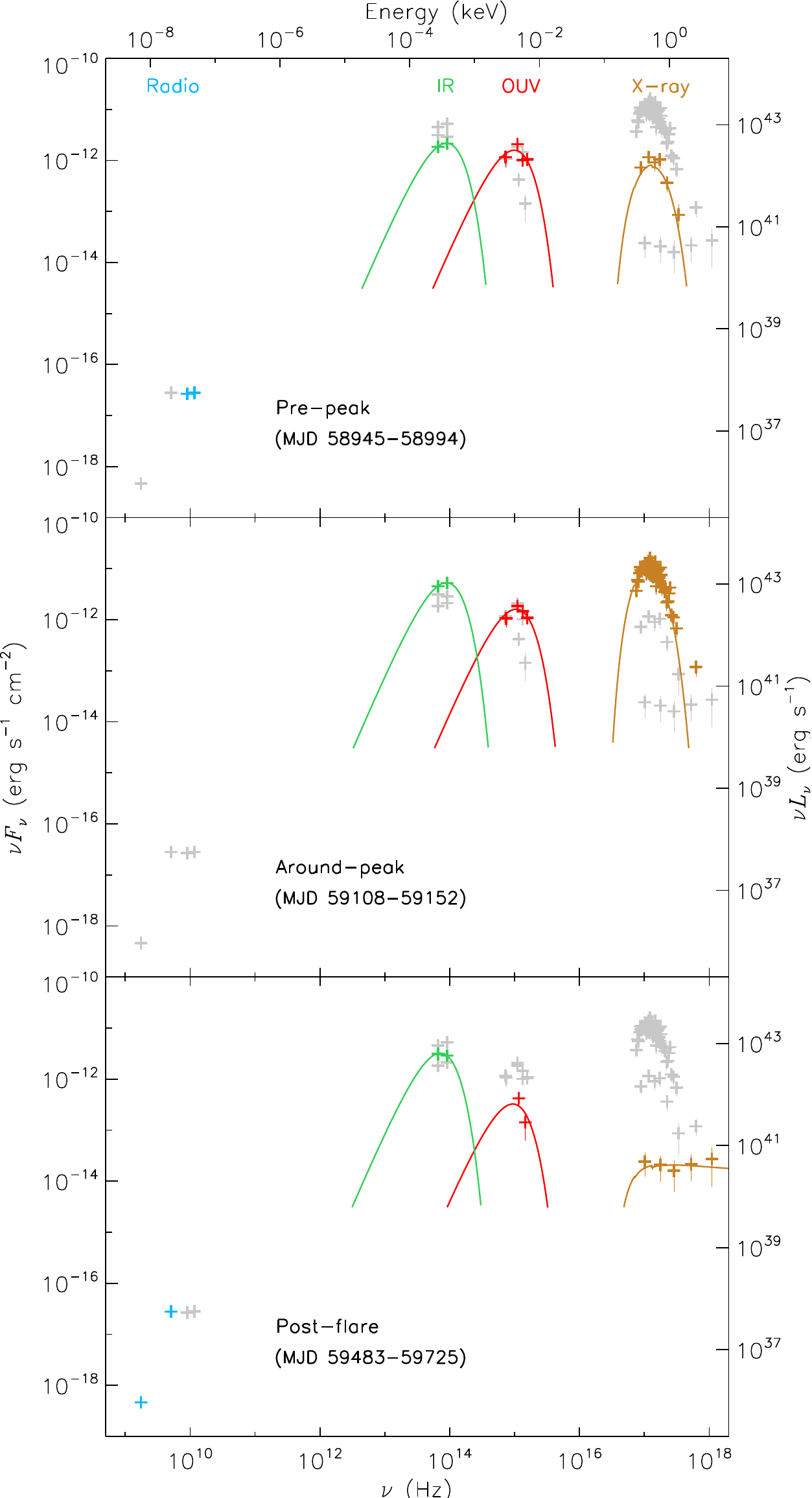}
\vspace{-0.3cm}
\caption{The radio to X-ray SEDs. The three SEDs are taken from the pre-peak, around-peak and post-flare periods (the peak and flare here are referred to the second optical flare, see legends for the time). The optical and X-ray spectra have been rebinned. The light blue, green, red and fawn crosses represent the radio, IR (flaring component), optical/UV and X-ray data, respectively. Data from different epochs are shown in grey for comparison. The radio SED shown in the bottom penal is taken 34--95\,d before the post-flare period.
}
\label{fig:avd_sed}
\end{figure}

\section{Spectral energy distribution}
\label{sec:sed}

We generate the spectral energy distribution (SED) of {\avd} from radio to X-ray bands with the VLA/VLBA, {\wise}, UVOT and XRT data.
The optical-to-UV spectrum is converted from the host-subtracted and Galactic extinction-corrected photometry in each UVOT filter (see more details in Appendix~\ref{sec:app}). 
As the optical emission in the V and B bands has approximately maintained at the host galaxy level, we exclude the V band data from the spectrum.
To better define the optical-to-UV spectrum, we only selected the UVOT observations with detection in at least four filters. However, no observations from the late-time flare (after MJD~59368) meet this criterion. We thus jointly fitted the optical-to-UV SED extracted from the observations between MJD~59368 and 59376 with all the parameters linked to increase the signal-to-noise ratio. The same method was applied to the X-ray spectra obtained in this period. The other ten selected UVOT spectra were fitted independently while the simultaneous XRT spectra were jointly fitted with the column density linked across observations. 
Here we only include the UVOT data for the spectral analysis; the detection provided by ZTF is likely dominated by line emission which we will discuss further in Sect.~\ref{sec:opt_UV}.
We illustrate three radio to X-ray SEDs ($\nu F_{\nu}$) in Fig.~\ref{fig:avd_sed}: from top to bottom they are taken from the pre-peak (MJD~58945--58994), the around-peak (MJD~59108--59152) and the post-flare (MJD~59483--59725) phases. To increase the signal-to-noise ratio, we have also combined the XRT spectra taken between MJD~59483 and 59725.



\begin{figure*} 
\centering  
\includegraphics[width=0.316\linewidth]{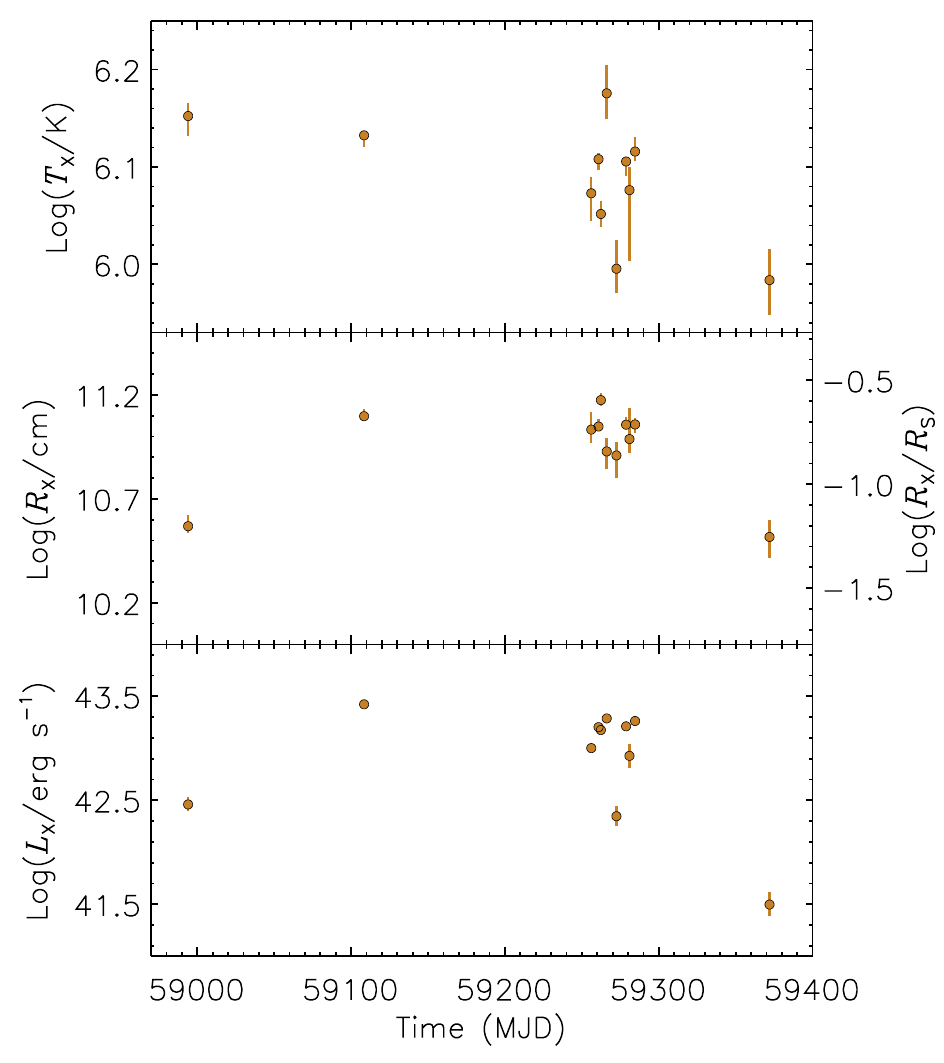} \hspace{0.0cm}
\includegraphics[width=0.316\linewidth]{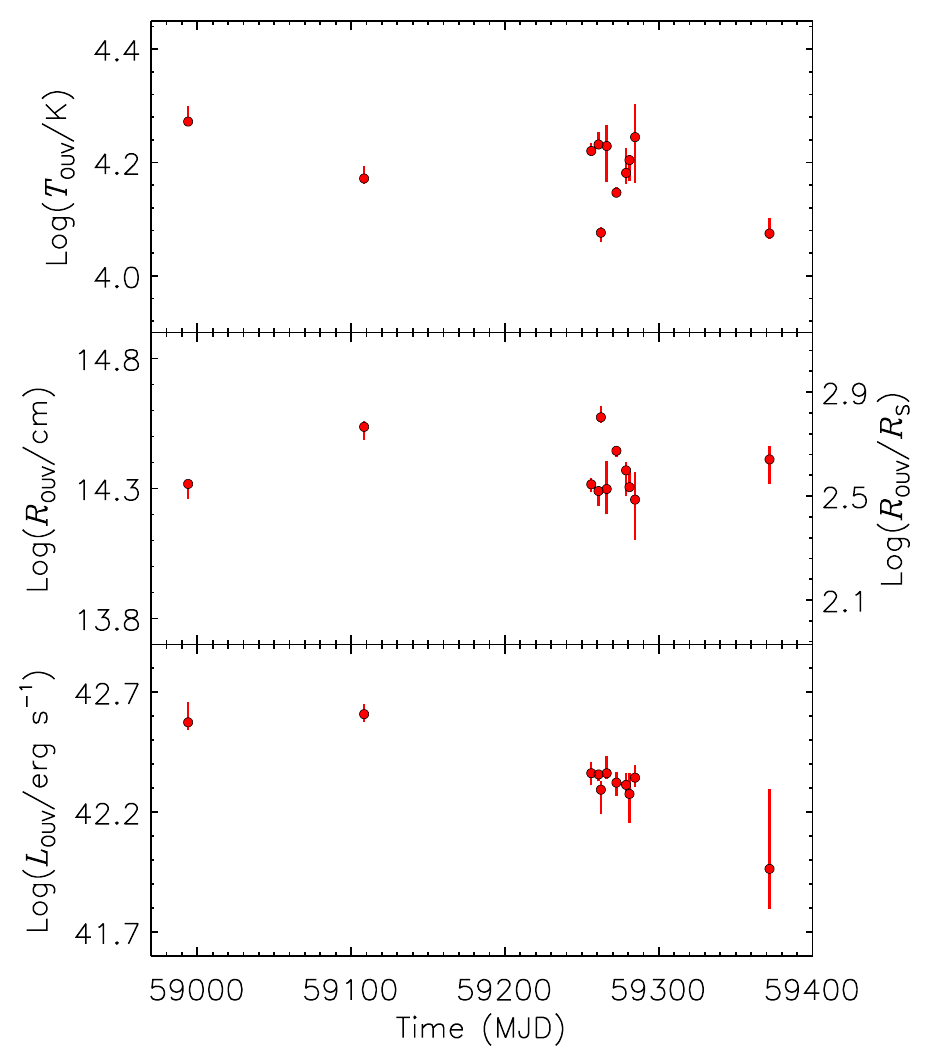} \hspace{0.cm}
\includegraphics[width=0.316\linewidth]{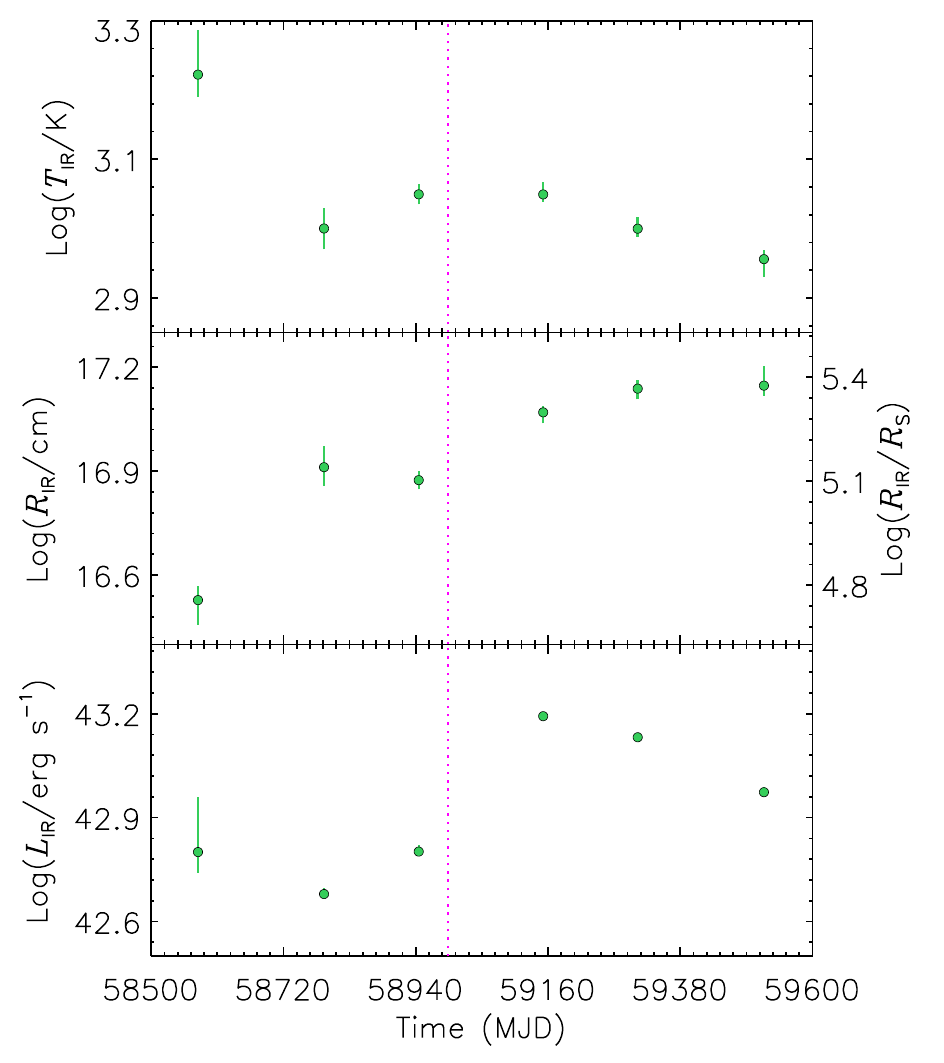}
\vspace{-0.2cm}
\caption{Blackbody temperature (top)/radius (middle)/luminosity (bottom) evolution of the X-ray (left), optical-UV (centre) and IR (right) emission. The magenta dashed line in the right panels indicates the position of the first data point shown in the other panels.}
\label{fig:spec_para}
\end{figure*}

Both of the optical/UV and the X-ray spectra can be described reasonably well with a(/an absorbed) single-temperature blackbody component (\texttt{bbodyrad} in \textsc{xspec}, \citealt{Arnaud1996}). The X-ray absorption (described by \texttt{tbabs}) is constrained to be larger than the Galactic value, $2.4\times 10^{20}\,\rm cm^{-2}$, adopted from the HI4PI survey \citep{HI4PI}. 
The blackbody radius, $R$, can be computed from the normalisation of the \texttt{bbodyrad} component, $K$, as $R=\sqrt{K}\times D_{10}$, where $D_{\rm 10}=1.3\times10^4$ is the source distance in units of 10\,kpc. 
Adopting a BH mass of $10^{6.3}\,M_{\odot}$ from \cite{Malyali2021}, we obtain the optical-UV and the X-ray photospheric radii $R_{\rm OUV}=611-1267\,R_{\rm S}$ and $R_{\rm X}=0.1-0.5\,R_{\rm S}$, respectively, where $R_{\rm S}=2GM_\mathrm{BH}/c^2$.
The optical/UV and X-ray fluxes are calculated in the 0.001--0.2\,keV and 0.3--10\,keV, respectively. We illustrate the evolution of the blackbody temperature, radius and luminosity in Fig.~\ref{fig:spec_para} and the best-fitting parameters inferred from the three SED from IR to soft X-ray in Table~\ref{fit_tab}.
Alternatively, we have also considered non-thermal origins for the optical-UV and X-ray emission. If we replaced the \texttt{bbodyrad} component with a \texttt{powerlaw} to fit the SED shown in the middle panel of Fig.~\ref{fig:avd_sed}, we obtained a photon index of $\Gamma_{\rm OUV}=2.2\pm0.3$ and $\Gamma_{\rm X}=6.5\pm0.3$, respectively. For the optical-to-UV SED, it results in a worse fit with $\chi^2$ increasing by $\Delta \chi^2=4.03$ for the same degree of freedom ($\nu=4$) and the null hypothesis probability decreasing from 0.02 to 0.007.
While for the case of the X-ray SED, we obtained a statistically good fit with an increase in the column density of nearly one order of magnitude. However, the photon index is too high to be non-thermal. Therefore, a thermal origin tends to be preferred by both the optical-UV and the X-ray SED. 
In fact, the measured temperature and radius are consistent with previous works in both optically and X-ray selected TDEs \citep{Gezari2021}.

Although with some small wiggles, neither the optical/UV nor the X-ray blackbody temperature shows statistically significant evolution, except for the one derived from one (MJD~59368--59376) of the combined spectra. Especially for the X-ray data, both of its temperature and photospheric radius are smaller than the rest period (see the last data point in leftmost panel of Fig.~\ref{fig:spec_para}).
This is likely due to the application of an inappropriate model. 
However, if replacing the \texttt{bbodyrad} component with a \texttt{powerlaw} component (\textsc{xspec} based) in our model, the two fits are comparable and the obtained photon index is $4.7\pm1.1$, still supporting a thermal spectrum. While as shown in the bottom panel of Fig.~\ref{fig:avd_sed}, the late-time (MJD~59483-59725) X-ray spectrum flattened with a photon index of $\Gamma=2.08_{-0.89}^{+1.53}$. Although the photon index can only be loosely constrained due to the low statistics, the spectrum at late times is apparently different from other periods. 
Hardening of the X-ray spectrum as the source evolved back to quiescence is also confirmed by the spectral analysis of \Ni data in Wang et al. (in prep). After MJD~59335, the {\Ni} spectrum only requires an absorbed \texttt{powerlaw} to account for the continuum, with the photon index decreasing from nearly 5 to $\sim$2 and the hardness ratio\footnote{The hardness ratio here is defined as the ratio between the count rates at 0.8--2.0\,keV and 0.3--0.8\,keV.} increasing from 0.1 to 1.
 
We show the spectral index in the IR band, $\alpha_{\rm IR}$, in the bottom panel of Fig.~\ref{fig:WISE}. For the steady component prior to the flare, $\alpha_{\rm IR} \approx 1.91\pm0.10$, which is typical of optically thick thermal emission (i.e. a black body spectrum with a temperature $> 3000$~K); during the first flare, $\alpha_{\rm IR}$ decreased from $0.9\pm0.2$ to being consistent with 0; in the second flare, $\alpha_{\rm IR}$ evolved to be negative.
By fitting the IR spectrum (only the flaring component) with a \texttt{bbodyrad} component we obtain a temperature decreasing from $1590\pm220$\,K to $873\pm35$\,K, and the corresponding photospheric radius increasing from $3.4\times10^{16}\rm \,cm$ to $14.0\times10^{16}\rm \,cm$ (see the right panels of Fig.~\ref{fig:spec_para}). In what follows, we only consider the flaring component in our analysis. We discuss why a non-thermal origin is not in favour in Sect.~\ref{sec:dust}.

In Fig.~\ref{fig:avd_sed} we also show radio data in the SED. 
These radio observations cover a long period of time ($\sim$1.5\,yr), so it is more prudent to compare the ones that are closer in time. 
The in-band 10-GHz VLA observation reveals an optically-thin regime, described by a steep $\alpha_{\rm 9-11.1 \,GHz}=-0.96\pm0.14$ ($F_{\nu}\propto \nu^{\alpha}$), while the VLBA observations -- which are 61\,d apart -- indicate that the source has an inverted spectrum in the 1.6--5\,GHz band, $\alpha_{\rm 1.6-5 \,GHz}= 1.02\pm0.20$. 
In order to elucidate the nature of \avd this novel information needs to be put in a temporal and multi-band context .

\section{Discussion}
{\avd} is a nuclear transient exhibiting two multi-wavelength flaring episodes. Based on the observations from radio to X-rays with VLA/VLBA, {\wise}, ZTF, {\sw}/UVOT, {\sw}/XRT and {\Ni}, it has been noted: (i) two consecutive optical flares (spanning over 1000\,d) showing different profiles -- the first one presents a sharp peak and the second one presents two plateaus; (ii) a compact radio core in the second flare ($L_\mathrm{10\,GHz} \sim 6 \times10^{37}\rm \,erg\,s^{-1}$) together with $L_\mathrm{bol} > 6.5\times10^{43}\rm \,erg\,s^{-1}$ at the peak; (iii) a $\sim$70-d delay between the optical and X-rays; (iv) two bright IR flares with luminosities up to $1.5\times10^{43}\rm \,erg\,s^{-1}$; (v) different temporal evolution in radio, IR, optical and X-ray bands from the second peak.
In this section we discuss each of these features and try to unveil the nature of this nuclear transient.

\subsection{The multi-band properties and possible origins}
\subsubsection{Radio}
\label{sec:radio}

Despite the uncertainties introduced by the non-simultaneity of multi-frequency radio observations and the lack of a high-cadence radio monitoring of the source, we can still derive insightful conclusions regarding the radio emission of {\avd}. First, we neglect any pre-existing radio activity of the galaxy prior to the detection of {\avd}, as previous observations in 2001 and 2017 did not show significant radio emission (though with an upper limit of $< 10^{38}$\,erg\,s$^{-1}$).

We estimate the brightness temperature $T_{\rm B}$ from the flux density of the compact component in the VLBA maps. We obtain $T_{\rm B} \sim 4.8\times 10^{6}$\,K at 5\,GHz and $T_{\rm B} \sim 1.7 \times10^{6}$\,K at 1.6\,GHz, corrected for the redshift of the source. These values are higher than the typical limit value, 10$^5$\,K, used for compact starburst \citep{condon91}. The high brightness temperature beyond this limit is generally interpreted as evidence of non-thermal synchrotron radiation from an accreting BH (e.g. \citealt{falcke00}).


Next, we examine the radio loudness, $R=L_{\rm radio}/L_{\rm disc}$, where $L_{\rm radio}$ is typically measured at 5\,GHz and $L_{\rm disc}$ is measured as the B-band luminosity (4400\,\AA, \citealt{kellermann89}) or the 2--10\,keV X-ray luminosity \citep{terashima03}. This parameter $R$ is used to distinguish between radio-quiet non-jetted and radio-loud jetted AGN \citep{padovani17}. Unfortunately, we cannot estimate a robust $R$ at the VLA detection because the X-ray spectrum is dominated by the background at energies above 2\,keV and the B-band flux is basically at the host galaxy level.
At the time of the 5\,GHz VLBA detection ($L_{\rm radio}\sim4.0\times 10^{37}\rm \,erg s^{-1}$), we opt for the 2--10\,keV X-ray luminosity to estimate $L_{\rm disc}$ since the X-ray spectrum hardens. In order to obtain a robust measurement, we jointly fitted the XRT spectra from MJD~59483 to 59725 (the X-ray flux remained constant in this period). The corresponding X-ray luminosity is $L_{\rm X}=(2.5$--$13.0)\times10^{40}\rm \,erg\,s^{-1}$, and hence the radio loudness is $\log{R_{\rm X}} = -(2.8$--$3.5)$, which is higher than the value of $-4.5$ used to distinguish between the radio-quiet and -loud regimes \citep{terashima03}. This indicates that the source was in a radio-loud state at the VLBA epoch with a radio power similar to those of nearby low-luminosity AGN (e.g. \citealt{nagar05,baldi21b}).

The flux densities derived from the 10-GHz VLA and 5-GHz VLBA observations differ by a factor $\sim$2, which could reconcile with an intrinsic radio variability of the source, assuming related to BH accretion (for typical radio-quiet AGN, \citealt{panessa19}). The VLA spectral index measured at the second optical flare, $\alpha_{\rm 9-11.1\,GHz}\sim -1$, significantly differs from the VLBA spectral index measured during the optical/X-ray luminosity drop (more than a year later), $\alpha_{\rm 1.6-5\,GHz}\sim 1$.
As our observations probed different frequency ranges at different times and physical scales, the possible interpretations on the steep-spectrum emission at higher frequencies and the inverted-spectrum emission at lower frequencies are not univocal and are discussed below:


\begin{figure} 
\centering  
\includegraphics[width=0.9\linewidth]{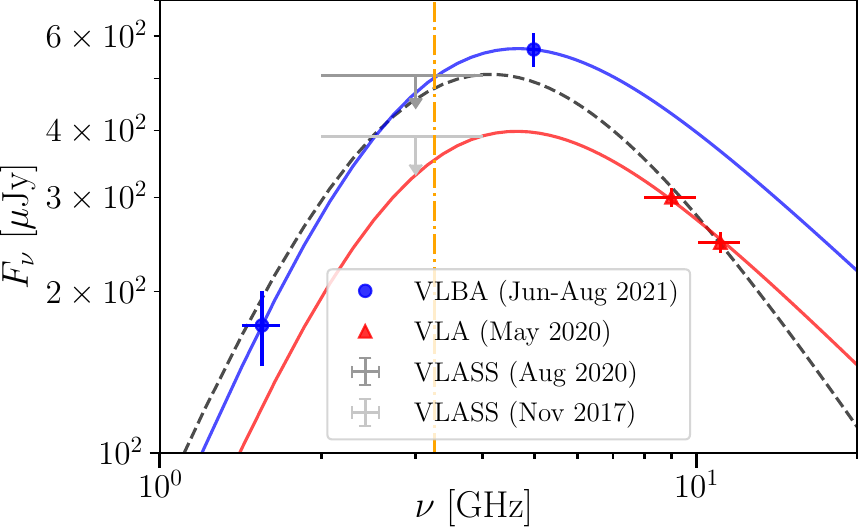}
\caption{Multi-epoch radio SED of \avd. The dashed black line is a fit to all the data using the blob model given by Eq.~\ref{eq:radio_SED}, while the solid red and blue lines are fits to only data taken with similar epochs and angular resolution (see text in Sect.~\ref{sec:radio}). This shows that the whole radio emission could be reconciled with a non-stationary synchrotron-emitting compact source with a flux increasing by 50\% from 2020 to 2021. The vertical dash-dotted line is the value $\nu_\mathrm{a}\approx 3.2$\,GHz at which synchrotron self-absorption starts to dominate the SED.}
\label{fig:radiofit}
\end{figure}



\begin{enumerate}
\item Stationary blob. In a first-order approximation, we can test the possibility that the whole radio emission of \avd is related to a single stationary synchrotron-emitting compact blob emitted by the accreting BH. The large caveat of this assumption is the omission of the effect of the different physical scales probed by the VLA and VLBA observations.
The synchrotron self-absorption frequency of the source, $\nu_\mathrm{a}$, marks the transition from an optically-thin synchrotron spectrum at high frequencies ($\alpha<0$ at $\nu \gg \nu_\mathrm{a}$) to a self-absorbed spectrum at low frequencies ($\alpha=2.5$ at $\nu \ll \nu_\mathrm{a}$). This behaviour is captured in the model developed by \citet{granot02} for synchrotron emission from gamma-ray burst afterglows. We use this model to fit the radio flux densities of \avd, similarly as done by \citet{cendes21} for the radio-TDE AT2019dsg.
Explicitly, this model parameterised the radio SED as\footnote{We consider that the lowest energy electrons emit at a characteristic frequency $\nu_m$ that is much lower than the frequencies of interest in our observations. Thus, Eq.~1 from \citet{cendes21} converts to Eq.~\ref{eq:radio_SED} for $\nu \ll \nu_m$.}:
\begin{equation} 
\label{eq:radio_SED}
F_\nu = F_0 \left( \frac{\nu}{\nu_0} \right)^{5/2} \left[ 1 + \left(\frac{\nu}{\nu_\mathrm{a}}\right)^{s_2 (b_2 - b_3)} \right]^{-1/s_2},
\end{equation}
where $F_0$ is the flux density at a reference frequency $\nu_0$, %
$b_2 = 5/2$, $b3 = (1 - p)/2$, with $p$ the spectral index of the relativistic electron population ($N(\gamma_\mathrm{e}) \propto \gamma_\mathrm{e}^{-p}$), and $s_2 = 1.25 - 0.18 p$. 
Figure~\ref{fig:radiofit} (dashed line) shows an exploratory fit to the whole dataset. This fit yields a very steep electron energy distribution with $p=3.8\pm0.7$ ($N(\gamma_\mathrm{e}) \sim \gamma_\mathrm{e}^{-p}$), which is steeper than typically found in TDEs \citep[though still physically possible under specific conditions; e.g. quasi-perpendicular shocks with velocities $> 10^4$\,km\,s$^{-1}$,][]{Xu2022}. Moreover, this fit cannot satisfactorily reproduce the radio data as it significantly underpredicts the flux density at 5\,GHz and it is also in tension with the VLASS 2017 upper limit at 3\,GHz. This test demonstrates that a single stationary blob is much unlikely a plausible scenario for \avd, because of its inconsistency with the observations.

\item Temporal evolution. We have also considered a more plausible physical picture in which the observed radio flux is produced by an evolving source. Based on the radio-AGN phenomenology \citep{panessa19}, two possible interpretations on the origin of the radio emission are considered:

\end{enumerate}

\begin{itemize}[leftmargin=1\parindent]

\item Evolving blob. In this scenario both the peak brightness of the blob and $\nu_\mathrm{a}$ change with time, with $\nu_\mathrm{a}$ shifting to lower frequencies (e.g. for blob expansion, \citealt{eftekhari18,cendes21}). 
The problem we face is that we only have two quasi-simultaneous data points at each epoch (2020 and 2021), and thus it is not possible to fit simultaneously all the parameters that describe the spectrum in Eq.~\ref{eq:radio_SED}. Thus, we assumed that the values of $p$ and $\nu_\mathrm{a}$ did not change significantly between the two observing epochs, and allowed only for the peak flux to vary.  
Fitting the high-frequency VLA observations we obtained $p \approx 2.9\pm0.4$, while the value of $\nu_\mathrm{a}$ is given by the low-frequency VLBA observations and the fitting yields $\nu_\mathrm{a} = 3.2\pm0.3$\,GHz. The high and low-frequency fits (blue and red lines in Fig.~\ref{fig:radiofit}) are consistent with a flux increasing by 50\% from 2020 to 2021. 
The results are shown in Fig.~\ref{fig:radiofit} and indicate that, in principle, an evolving single emitting blob can account for the behaviour seen in the different radio bands at different epochs. However, this requires that the value of $\nu_\mathrm{a}$ did not change significantly between 2020 and 2021, as a value of $\nu_\mathrm{a} > 4$\,GHz in 2020 \citep[as could be expected if it decreases in timescales of hundreds of days, e.g.][]{cendes21} would compromise the fitting of the VLA spectra. A rather steady value of $\nu_\mathrm{a}$ has been observed in other radio TDEs \citep{cendes22,goodwin22} and stands opposite to the expectations for a cooling/expanding single blob. A possible explanation is the ejection of consecutive radio-emitting blobs that prevent $\nu_\mathrm{a}$ from shifting to lower values. To test this scenario more convincingly, a multi-frequency radio monitoring at high angular resolution is needed.


\item Wind-jet scenario. Alternatively, the radio spectrum of {\avd} can also be phenomenologically interpreted with two different physical processes operating in two separate epochs. First, during the VLA detection (coincident with the optical/X-ray flare) when the accretion disc is formed, the radio properties are consistent with an optically thin large-scale outflow (with a speed a few thousands of km\,s$^{-1}$, \citealt{zakamska14}), driven by the radiation pressure of the accretion disc. This disc wind shocks the surrounding gas and accelerates relativistic electrons that produce synchrotron radio emission on scales $>$ 100\,pc \citep{zakamska14,nims15,karouzos16}. According to the wind model from \cite{nims15}, the synchrotron emission from AGN-driven winds has $\alpha = -1$ and a luminosity $L_{\rm radio} \sim (0.01$--$1)\times 10^{-5} \, L_{\rm bol}$\footnote{The range depends on the fraction of the shock energy which goes
into relativistic electrons.}, where in our case $L_{\rm bol} \sim L_{\rm Edd} \approx 3\times10^{44}$~erg\,s$^{-1}$ (although model-dependent, see Sect.~\ref{sec:xray}), yielding $L_{\rm radio} \sim (0.01$--$1)\times10^{39}$~erg\,s$^{-1}$. Such a range of values include the radio luminosity measured at 10\,GHz ($\sim 6 \times 10^{37}$~erg\,s$^{-1}$) if the source is accreting at $\dot{M} \gtrsim 0.1 \dot{M}_{\rm Edd}$. Furthermore, the high Eddington ratio (greater than unity, see Sect.~\ref{sec:xray}) and the VLA spectral index of AT~2019avd lie on the empirical relation found for AGN-driven outflows \citep{laor19,yang20}. 
At later epochs, a jet would be responsible of the emission detected during the VLBA observations. In fact, the evidence -- i.e. the inverted radio SED, the moderately high $T_{\rm B}$, the pc-scale compactness, and the radio-loud classification -- indicates that the radio emission is not due to star formation, but rather to a newly formed compact radio jet ($< 10$~pc, based on the VLBA map), in analogy with young radio sources \citep[e.g.][]{odea21}. A mildly-to-low relativistic jet (probably with a jet bulk Lorentz factor $<$ 2) is favoured over i) a slow outflow, since the latter is expected to have a diffuse morphology (rather than compact on pc-scales) and an optically-thin spectrum \citep{falcke95,panessa19}, and ii) a highly-relativistic jet because $T_{\rm B}$ of \avd is lower than typical values measured from radio-loud AGN, $T_{\rm B} \sim 10^{8}$--10$^{13}$\,K (e.g. \citealt{ghisellini93,nagar05}), with a typical jet bulk Lorentz factor $\gtrsim$ 2--3 \citep{urry91,urry95,mullin09,hogan11}. 
The contribution to the radio emission from the disc wind at the VLBA epoch is not expected to be important as the mass accretion rate has decreased to the sub-Eddington regime, resulting in a strong decline of the outflow, although the unknown physical conditions of the disc and surrounding gas (e.g. magnetic field, density and temperature) prevent from deriving a more conclusive interpretation.  


\end{itemize}



\subsubsection{Infrared}\label{sec:dust}
IR emission can, in principle, originate from either non-thermal or thermal processes. In the former case, if we associate the IR emission to optically-thin synchrotron radiation, the radio emission from this component would be much brighter than the observed one. Namely, the radio emission at 10\,GHz would need to be $\sim$10\,mJy, which is roughly 33 times higher than our 10-GHz VLA detection. This discrepancy rules out a non-thermal origin for the IR activity of {\avd}. We refer the reader to Appendix~\ref{sec:non_thermal_IR} for further details of this calculation.

Alternatively, IR emission has been commonly taken as the evidence of dust echos in both AGNs \citep{netzer15} and TDEs \citep{Lu2016,Jiang2021,Velzen2021}, i.e. optical, UV and X-ray photons are absorbed by dust grains and re-radiated in the IR. In the case of \avd, the IR emission is very bright, reaching up to $1.5\times10^{43}\rm \,erg\,s^{-1}$, and is also significantly variable, having increased twice during the two flares (by an additional factor 2--3 during the second one with respect to the first one).
The slight increase in the IR temperature by $\Delta T_{\rm IR} \sim 120$\,K (with a confidence of $12\sigma$) from the first to the second flare also supports the presence of the second IR flare (see Fig.~\ref{fig:spec_para}). 
The first flare can be explained by the dust reprocessing the optical/UV light, while the second one could be associated with the soft X-rays (and/or EUV). Since the optical/UV fluxes in the two flares are comparable and the X-ray is much brighter than the optical/UV in the second flare, the further increase in the IR is very likely due to the additional contribution from the X-rays. 
This hints that there might be no strong X-ray/EUV emission in the first flare.
The lack of Bowen features in the first flare \citep{Malyali2021} could be indicative of the absence of X-ray/EUV emission \citep{Leloudas2019}.

One important application of dust echos is to calculate the bolometric luminosity\footnote{Since several bands are unobserved (e.g. EUV) and the dust would obscure the optical/UV/X-ray emission, the actual bolometric luminosity should be higher than the one we estimate. \label{footnote:bol_lumi}}. 
In the context of TDEs, \cite{Velzen2016} assumes that the reprocessing shell of dust is located at the dust sublimation radius given by
\begin{equation}
    R_{\rm dust} \approx  0.15\, \left(\frac{L_{\rm 45}}{a_{0.1}^2 T_{\rm 1850}^{5.8}}\right)^{1/2} \, {\rm pc},
\end{equation}
where $L_{\rm 45}$ is the absorbed luminosity in units of $10^{45}\rm \,erg\,s^{-1}$, $a_{0.1}$ is the size of the dust grains in units of $0.1\rm \, \mu m$, and $T_{1850}$ is the dust temperature in units of 1850\,K. 
Another parameter required to measure $L_\mathrm{bol}$ is the time delay of the dust echo with respect to the TDE flare, $\tau \sim R_{\rm dust}/c$ \citep{Jiang2016}. 
If we assume that the first IR data point (MJD~58578) of the flaring component is the peak of the first IR flare, we can get a time delay of $\tau \sim 40$\,d with respect to the first optical flare. This leads to $R_{\rm dust}\sim 0.03$\,pc, which is comparable to the photospheric disc radius of 0.01~pc (considering that the latter is likely underestimated).
Adopting a single dust radius of $a=0.1\rm \,\mu m$, we obtain an absorbed bolometric luminosity of the dust of $\sim 0.24\,L_{45}$ (or $\sim 0.94\,L_{\rm Edd}$). This implies a covering factor of the dust in the nuclear region of \avd, $f_{\rm dust} \sim L_{\rm IR}/L_{\rm abs} \sim 0.1$. Considering the uncertainties in both the time lag and the size of the dust grains, these values, i.e. $L_{\rm abs}$ or $L_{\rm bol}$ and $f_{\rm dust}$, should be taken with caution.
While in terms of the second flare, the case is more complicated because of the overlap of the dust cooling process (due to the first flare) and the extra heating from the second flare, as well as the involvement of the soft X-rays (which are highly variable and peaked at least twice) and/or the unobserved EUV. At the time of writing, the dust emission is still in the cooling phase, so a further, comprehensive investigation of the IR activity in {\avd} will be conducted in a separate paper.

\begin{figure}
\resizebox{0.8\columnwidth}{!}{\rotatebox{0}{\includegraphics[clip]{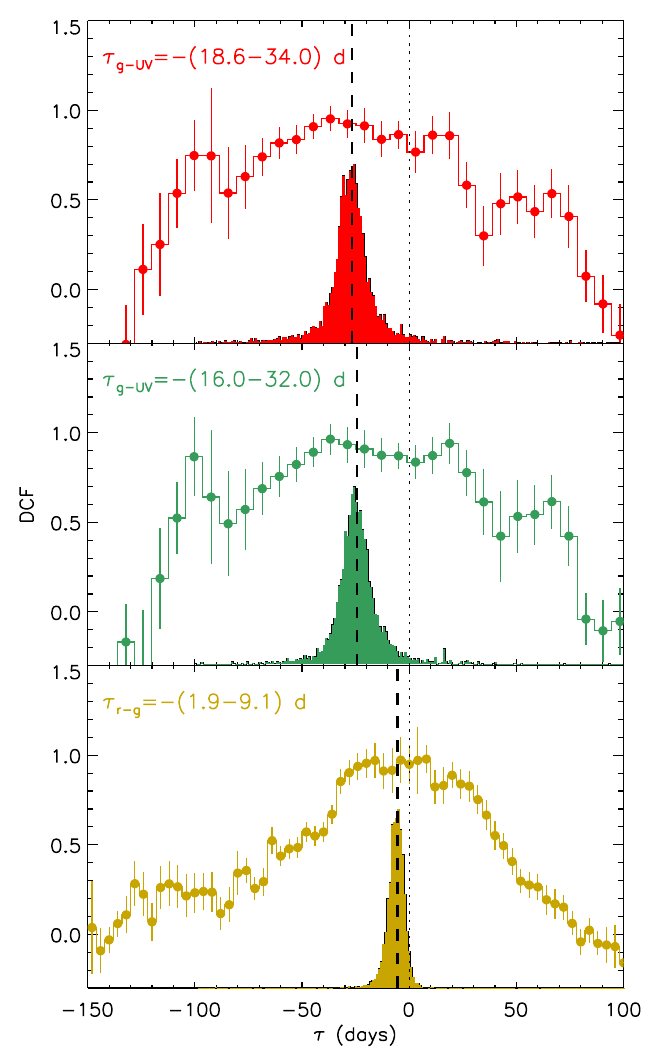}}}
\vspace{-0.2cm}
\caption{Inter-band lag measurements. The contours represent the bootstrap distribution which has been re-scaled for plotting purpose. The dashed lines indicate the median of each distribution.}
\label{fig:lag}
\end{figure}

\subsubsection{Optical/UV}\label{sec:opt_UV}
The optical/UV lightcurve in {\avd} has shown a peculiar pattern. The total optical emission lasts for over 1000\,d and shows two distinguished flares: the first one displayed a sharp peak around MJD~58534.3; the second one lasted roughly twice as long as the first one with a plateau phase present from MJD~59110 to 59220. 
Moreover, the first flare declined much faster than the second one, e.g. 74\,d after the first peak, the luminosity dropped by over a factor of 4, while the second flare only dropped by a factor of 2 in the same time.
We lack UV observations during the first flare and therefore cannot assess its behaviour during that time, but at least during the second flare the UV behaviour is comparable to the optical one. The following late-time (after the ZTF monitor) UV emission shows there was a second plateau present at least from MJD~59473 to 59580.

For the optical emission, we notice that the second flare has only been significantly detected by ZTF, but not by UVOT with its V and B bands. The lower panel of Fig.~\ref{figA:UVOT_ph} illustrates that the variability of the flare increase as the wavelength decreases. The magnitudes of the UVOT-V and -B bands remain nearly constant and are consistent with the host galaxy emission, indicating that the optical emission of {\avd} in these two bands is comparable to or lower than the emission from the host galaxy.
As the maximum effective temperature of the accreting material is proportional to the mass accretion rate and inversely proportional to the central BH mass \citep{Shakura1973}, the fact of the optical-UV SED of \avd dominated by the UV band suggests that its central SMBH has a small mass. A similar UV dominated SED \citep{Leighly2004} has been observed in the narrow-line Seyfert 1 Galaxy 1H~0707--495 which has a BH mass of $5\times10^{6}\,M_{\odot}$ \citep{Zoghbi2010}. 
The main difference between the ZTF-$g$ and -$r$ bands and the UVOT-V and -B bands is that the former are more sensitive to the H$\alpha$ and H$\beta$ line emission. Moreover, \cite{Malyali2021} observed significant, broad H$\alpha$ and H$\beta$ emission lines with a single-peaked profile in the optical spectra of {\avd}. Therefore, the optical flares observed by ZTF could be possibly contributed by the Balmer lines.

As the UV emission shows a comparable evolution to the optical line emission and the former shows higher variations than the latter, the second optical flare observed by ZTF is likely produced by UV reprocessing. 
To verify this conjecture, we applied the same method described in Sect.~\ref{sec:multiwavelength} (DCF plus bootstrapping technique) to estimate the lag between the optical and the UV. 
We obtained values of $\tau_{\rm r-UV}=-26.7_{-7.3}^{+8.1}$\,d between ZTF-r and UVOT-UVW1, and $\tau_{\rm g-UV}=-24.5_{-7.5}^{+8.5}$\,d between ZTF-g and UVOT-UVW1. 
We plot the DCF against lag in Fig.~\ref{fig:lag}, accompanied with the bootstrap distribution.
Such a negative lag between the emission at long and short wavelength, accompanied by the comparable behaviours, has been commonly taken as evidence of a reprocessing effect (e.g. reverberation mapping in AGN and XRBs; \citealt{Blandford1982,Peterson2004,Uttley2014}). 
The measured lags between the optical and the UV support the above conjecture. Additionally, if applying the same technique to the ZTF-r and -g bands, we obtained a marginally soft lag of $\tau_{\rm r-g}=-5.5\pm3.6$\,d. 
Considering only the light-travel time, the timescale of the lags between the optical and UV corresponds to a distance of $4.1-8.6\times10^{16}\rm\,cm$. This implies that the distance between the optical emitting region and the central SMBH is more than three orders of magnitude larger than the circularised debris disc ($R_{\rm circ}=2R_{\rm T}$, where $R_{\rm T}=1.8\times10^{13}\, \rm cm$ is the tidal radius, \citealt{Hills1975,MacLeod2012}) if assuming a SMBH with a mass of $10^{6.3}\,M_{\odot}$ disrupts a sun-like star. Therefore, the reflecting surface cannot be the disc itself, but some distant gas instead.
Furthermore, these soft lags are roughly a factor of 2 smaller than the dust-echo timescale (see Sect.~\ref{sec:dust}), suggesting that the gas reprocessing region locates closer to the centre than the dust reprocessing region, but they are adjacent to each other.
Overall, we conclude that the sub-pc scale environment of the circumgalactic medium of \avd is gaseous and dusty.

Additionally, by fitting the SED from the second flare with a single-colour blackbody, we obtained the photospheric radius of the UV/optical region of $R_{\rm OUV}\sim1.8-3.7\times10^{14}\rm \,cm$, which is about three orders of magnitude larger than the X-ray region, $R_{\rm X}\sim 0.3-1.5\times 10^{11}\rm \,cm$ (see Fig.~\ref{fig:spec_para}). This suggests the two emission regions are separated. Although the origin is still controversial, this phenomenon has been commonly observed in TDEs (see Fig.~8 in \citealt{Gezari2021}).


\begin{figure}
\includegraphics[width=\linewidth]{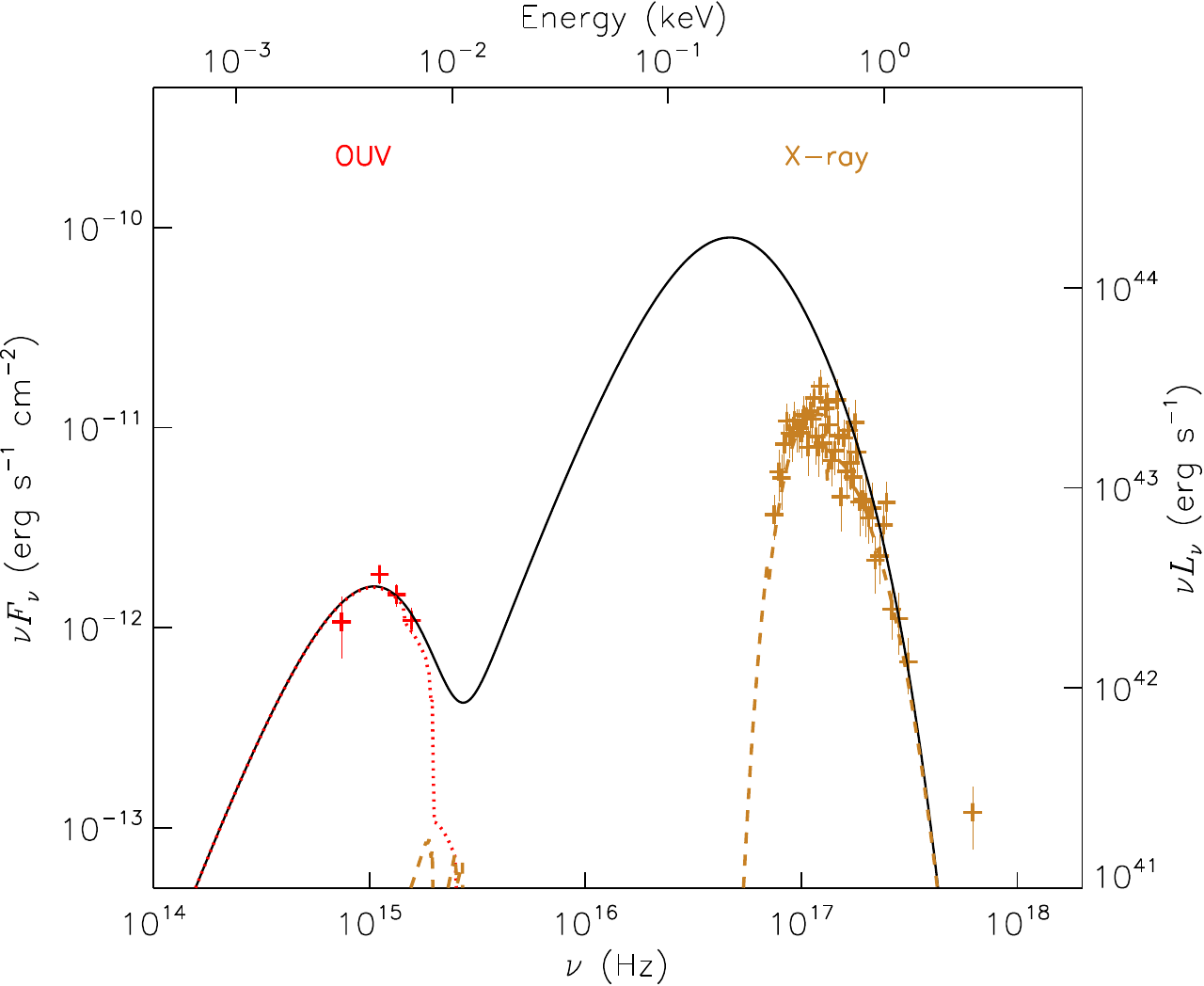}
\vspace{-0.3cm}
\caption{Optical to X-ray SED fitted with a \texttt{bbodyrad} (red dotted line) plus an absorbed \texttt{agnsed} component (fawn dashed line). The red and fawn crosses and lines represent the optical-UV and X-ray data. The black solid line represents the unabsorbed model. This SED is the same as the one shown in the middle panel of Fig.~\ref{fig:avd_sed}.}
\label{fig:mo_agnsed}
\end{figure}

\subsubsection{X-ray}
\label{sec:xray}
The further increase in the IR luminosity and the emergence of the Bowen line \citep{Malyali2021} associated with the second flare suggest that the soft X-ray emitter might not have been formed until the second flare.
Hence, a high accretion rate is required to efficiently form an accretion disc in roughly 100\,d. If simply adopting $\dot{m}=\dot{M}/\dot{M}_{\rm Edd}=L/L_{\rm Edd}$, the accretion rate will be underestimated (see the previous footnote~\ref{footnote:bol_lumi}). To obtain a reliable accretion rate, we employ an absorbed AGN SED model \texttt{agnsed}\footnote{\url{https://heasarc.gsfc.nasa.gov/xanadu/xspec/manual/node132.html}} \citep{Kubota2018} to describe our SEDs. However, the UV part of the SED cannot be described well with one \texttt{agnsed} component, for which an additional component is required. After adding a \texttt{bbodyrad} component to the model, the fit significantly improved. For instance, the reduced $\chi^2$ of the fit\footnote{For the setup of the \texttt{agnsed} component, we assume a non-spin SMBH, a relatively low disc inclination of $35\degree$, and adopt a mass of $10^{6.3}\,M_{\odot}$ \citep{Malyali2021}, a redshift of 0.028 and a distance of 130\,Mpc. As there is no hard X-ray emission, we assume that the radius, electron temperature and photon index of the hot Componization region are $6\,R_{\rm g}$, 100\,keV and 2, respectively. The radius of the warm Comptonisation region is $500\,R_{\rm g}$, the scale height of the radius is $10\,R_{\rm g}$ and no reprocessing is considered. Overall, besides the accretion rate and the outer radius, we also obtain a steep photon index and a low electron temperature for the warm Componization region, which are $\sim3.31$ and $\sim0.15$\,keV.} to the SED (see the middle panel of Fig.~\ref{fig:avd_sed}) around the X-ray peak decreased from 136.60/80 to 81.99/78. 
Fig.~\ref{fig:mo_agnsed} shows the unfolded spectra with the two components.
The best-fitting parameters of the \texttt{bbodyrad} component are comparable to the second data points in the left panels of Fig.~\ref{fig:spec_para}.
The inferred accretion rate is $6.7_{-2.2}^{+2.7}\,\dot{M}_{\rm Edd}$ and the outer region of the disc locates at $5.3_{-0.6}^{+0.9}\,R_{\rm S}$. Such a compact and soft X-rays dominated disc is very different from AGN's but is consistent with TDEs' \citep{Rees1988}. This result yields that the disc was in a super-Eddington regime, $\sim6.7 \,L_{\rm Edd}$, around the peak of the second flare.

In addition, the luminosity/accretion rate of {\avd} may be (further) underestimated if outflows are present.
The soft X-rays in {\avd} show the highest variability across all the electromagnetic waves, e.g. $L_{\rm X}$ varying a factor of 4 in 10\,d. Such a variability seems to be correlated with the luminosity: the variability was relatively low prior to the X-ray peak (at around MJD~59000), and it reduced significantly after the X-ray luminosity dropped below $5\times 10^{42}{\rm \,erg\,s^{-1}} \sim 0.02\,L_{\rm Edd}$ (Fig.~\ref{fig:lc}). This implies a state transition occurring around this X-ray luminosity. Although without direct evidence, outflows have been commonly observed in accreting systems \citep{Diaz-Trigo2016,Tetarenko2018}, in TDEs \citep{Dai2018,Hung2019,Hung2021} and in AGN \citep{king15,fiore17}. The X-ray variability of {\avd} could be connected with the obscuration of clumpy outflows, suppressed after a decrease of the luminosity/accretion rate (meanwhile the variability decreased too). This also means that the variation in the X-ray radius shown in Fig.~\ref{fig:spec_para} could be unphysical as the changes in the X-ray luminosity are not purely linked to the motion of the inner region of the accretion disk. Further analyses of the X-ray variability in \avd will be presented in Wang et al. (in prep.). Additionally, \cite{Chen2022} interpreted the large-amplitude X-ray variability as the rigid-body precession of the misaligned accretion disk caused by the Lense–Thirring effect of a spinning SMBH.

Additionally, in the late-time of the second flare, when the optical and X-ray luminosities decrease, the X-ray spectrum hardens with emission above 2\,keV (although very faint, see Fig.~\ref{fig:avd_sed}). This probably suggests the formation of a corona, which could lead to the formation/ejection of a compact jet. In fact, the corona has been interpreted as the base of a jet in AGN (e.g. \citealt{markoff05,king17}). The formation of an X-ray emitting corona is also temporally associated with the evolution into an accretion disc with a very low Eddington ratio ($\sim$ 10$^{-3}$), consistent with radiatively-inefficient accretion disc (e.g. \citealt{yuan14}). The sub-Eddington disc, the X-ray corona and the launch of a jet all reconcile with the typical picture of radio-loud jet-mode AGN (e.g. \citealt{falcke04,heckman14,hardcastle17}) and TDEs (e.g. \citealt{Velzen2011,Tchekhovskoy2014}). Recently, \citet{sfaradi22} interpreted the late-time radio emission of the TDE AT~2019azh  as result of accretion transition to a low hard state similar to the observed behaviour in BH X-ray binaries \citep{fender04}.


As mentioned in Sect.~\ref{sec:sed}, the disc radius inferred from a single blackbody model is about $0.1-0.5\rm \,R_{\rm S}$, smaller than the plausible event horizon of the central BH. A similar issue has been reported in previous works in the TDE literature, e.g. \cite{Wevers2019,Hinkle2021,Cannizzaro2021} and has been related to the underestimation of the X-ray luminosity. Several explanations have been proposed, such as disc inclination angle effects \citep{Stein2021}, obscuration \citep{Mummery2021} and/or Comptonisation \citep{Saxton2020} of disc photons. Given the limited evidence available to us, we cannot draw further conclusions on this.

\begin{figure*}
\resizebox{1.3\columnwidth}{!}{\rotatebox{0}{\includegraphics[clip]{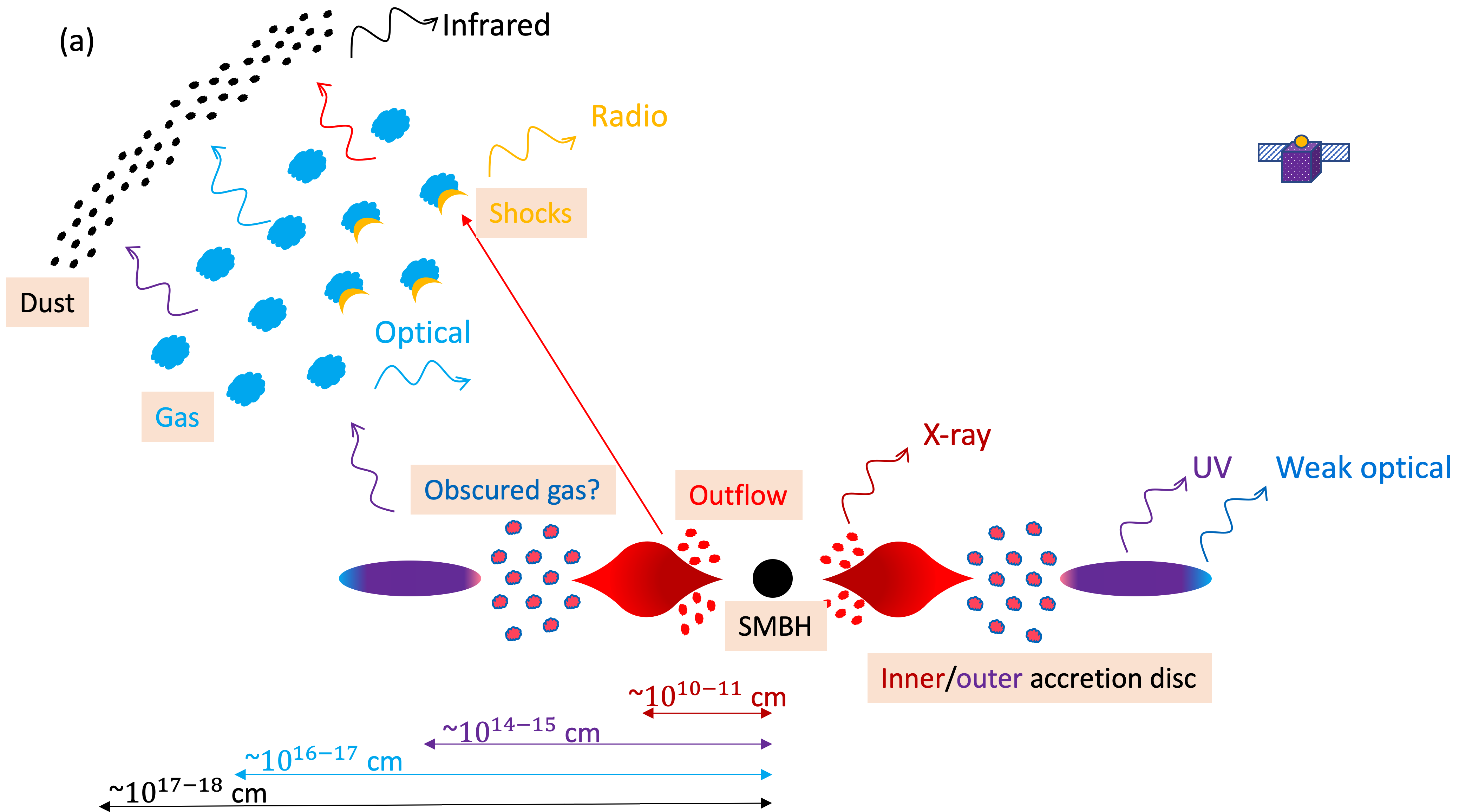}}}
\vspace{1cm}

\resizebox{1.3\columnwidth}{!}{\rotatebox{0}{\includegraphics[clip]{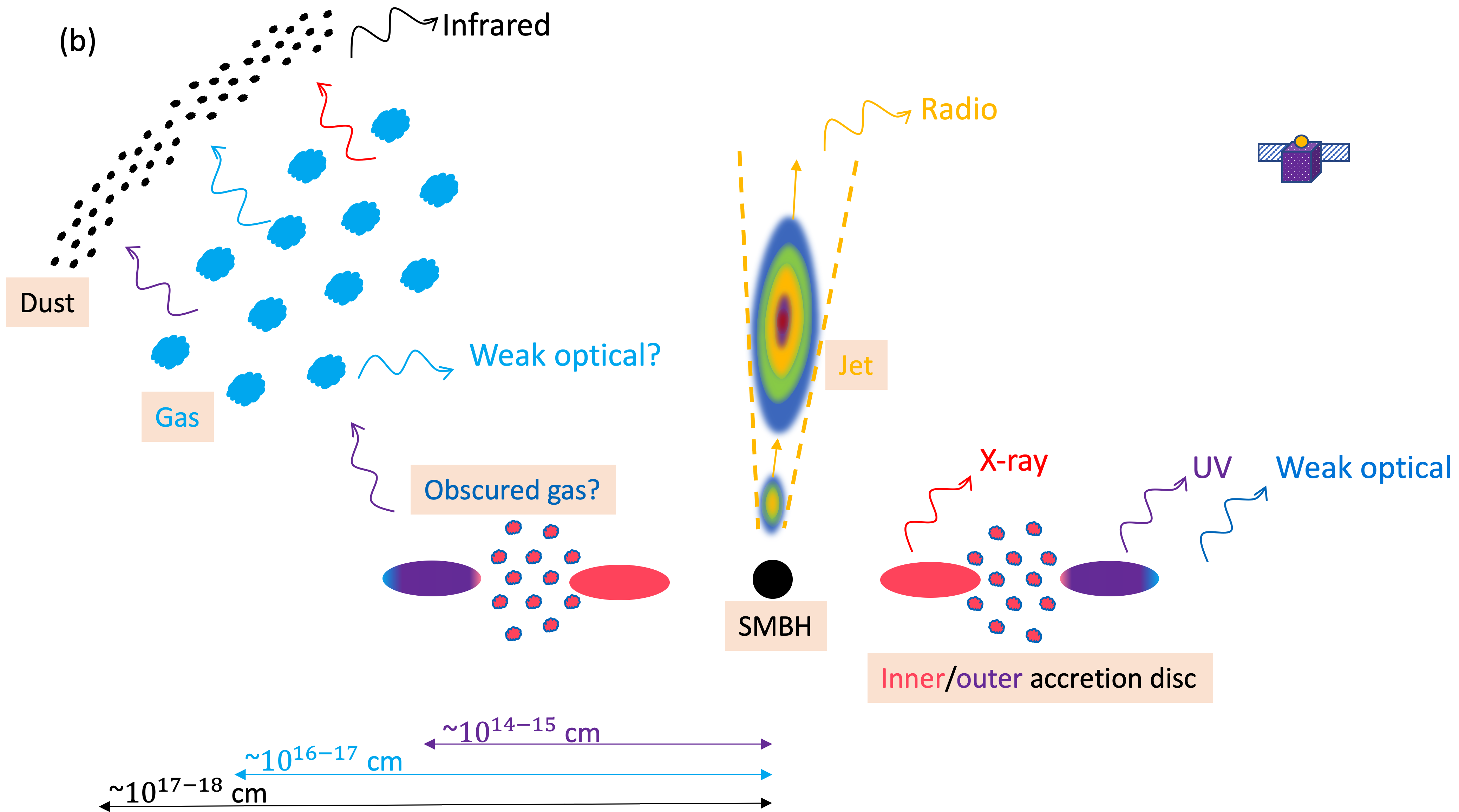}}}
\caption{Schematics of the multiwavelength emission evolving from the super-Eddington (a) to the sub-Eddington (b) regimes. The main changes between the two regimes are i) the inner accretion disc evolves from a puffy to a thin disc geometry; ii) the radio emission was first from the shocks between the outflows and distant gas and then from the late-time launched jets (an alternative scenario for the radio emission is discussed in \ref{sec:radio}). To simplify the schematics, the gases which are shocking with the outflows and are reprocessing the UV emission are not separated. The scales of each emitting region are not directly observed but rather inferred either from their luminosity or from the time lag.}
\label{fig:schemetic}
\end{figure*}

We show two schematics of the overall multiwavelength emission in super- and sub- Eddington regimes in Fig.~\ref{fig:schemetic} and discuss the possible nature of {\avd} based on the proposed geometry in the next section.

\subsection{Scenarios of the delay between optical/UV and X-rays}\label{sec:lag}
Thanks to the high cadence of ZTF and {\Ni}, we are able to closely monitor the second flare of {\avd} in the optical and X-ray bands. 
As shown in the long-term lightcurves (see the upper panel of Fig.~\ref{fig:lc}), after an (at least) 106-d plateau phase, the optical flare started to decline $\sim70$\,d earlier than the X-ray. This time delay has also been measured from the DCF between the optical (ZTF-$g$ and $r$ band) and the X-ray (\Ni) data, indicating that the optical decay leads the X-ray component in {\avd}. 
Due to the limited cadence of \sw, we cannot directly measure the delay between the UV and the X-rays. However, as reported in Sect.~\ref{sec:opt_UV}, the UV arrived the observer $16-34$\,d earlier than the optical, which means that the delay between the UV and the X-ray decay should be larger, roughly 90--130\,d.


A long hard lag can be interpreted in the framework of viscous propagation of the optical/UV fluctuations across the disc \citep{Lyubarskii1997,Marshall2008}. The viscous timescale significantly depends on the height-to-radius ratio, $t_{\rm visc} \propto (H/R)^{-2}$, where $H$ and $R$ are the vertical scale height and the radius of the disc, respectively.
Assuming the X-ray and the optical-UV photospheric radii (see Fig.~\ref{fig:spec_para}) as the radius of the inner and the outer part of a Shakura-Sunyaev disc \citep{Shakura1973} surrounding a SMBH with a mass of $10^{6.3}\,M_{\odot}$, the viscous timescale would be of several thousand of years, which is much larger than the observed lag. However, when the accretion rate is super-Eddington, the disc would be with a geometrically thick structure, similar to a slim disc \citep{Abramowicz1988} and thus the height-to-radius ratio $H/R$ would tend to unity. Consequently, the viscous timescale would be significantly reduced, which in turn could be consistent with the observed delay.

An alternative scenario to explain a delay between optical and X-rays has been proposed by \cite{Gezari2017} in the context of TDEs. They observed that in the TDE ASASSN--15oi the X-ray emission increased while the optical/UV flare declined. They thus proposed that the optical/UV originated from the self-interaction of the debris stream and the X-ray could be coming from delayed accretion through a newly formed debris disc. 
To discuss our result in this scenario, we need to consider whether the second flare in \avd could be due to stream--stream shocks.
As seen from Fig.~\ref{fig:lc}, the second flare started after the first flare nearly quenched. Therefore, even if the first flare is caused by stream--stream shocks, the rising of the second flare must be due to another process. This flare is unlikely triggered by another TDE given that its profile is inconsistent with a typical tidal disruption flare. 
Even though a plateau has been observed in the optical/UV photometry of some TDEs \citep[e.g.][]{Gezari2017,Wevers2019}, they are usually interpreted as an additional contribution due to X-ray reprocessing. 
After the optical/UV emissions start to decline, a plateau can only be formed if at the same time the X-ray emission increases (as the obscuring/reprocessing materials can only reduce). However, in the case of {\avd}, while a plateau appeared in the optical/UV photometries, the X-ray radiation reduced by nearly an order of magnitude (MJD~59110--59220). Therefore, the behaviour of the optical/UV emission in the second flare cannot be simply explained by circularization shocks plus an X-ray reprocessing process. We thus exclude this scenario as an explanation for the delay.

\subsection{Nature of the nuclear transient}
\label{sec:nature}
Here we discuss the nature of the transient \avd based on the origin of its multi-band emission. 
 
Although rarely observed, a SN can occur in the nuclear region of a galaxy (e.g. \citealt{Ulvestad1997,Mattila2001,Villarroel2017}). If only considering the post-peak period of the second optical flare of {\avd}, the lightcurve with a plateau lasting for over 106\,d is alike to that of a type II-P supernova (e.g. \citealt{Filippenko1997,Pastorello2004}). In such a system, the plateau phase has been suggested to be mainly powered by hydrogen recombination and radioactive decay \citep{Woosley1987}. However, the X-ray spectrum of SNe is usually hard, which can extend up to 100\,keV \citep{Ofek2013}. Furthermore, the emission in the optical is thermal, and the temperature of the ejecta is expected to increase more significantly along with the drop in luminosity.
These properties of SNe are incompatible with the ultrasoft X-ray emission, the nearly constant optical-to-UV temperature and the high brightness temperature of the detected compact radio source of {\avd}. 

Although the radio properties cannot be unambiguously interpreted, the most realistic scenario to account simultaneously for the radio and multi-band properties of {\avd} is a temporal evolution of an accreting BH, whose outflows produce non-thermal emission.
This can be in line with two plausible natures of the nuclear transient: an AGN-like phenomenon or a TDE.

In the former case, the non-detection of the active BH prior to its outburst in 2020 does not preclude the activity of central object, but sets an upper limit on its luminosity of $<$10$^{41}$\,erg\,s$^{-1}$ in X-rays and $<$10$^{38}$\,erg\,s$^{-1}$ in radio. These values are within the typical range of luminosities of nearby low-power AGN (e.g. \citealt{nagar05,balmaverde06,ho08}). The outburst could be then interpreted as due to an increase of BH accretion. The peculiar spectroscopic properties of the transient identify two possible AGN classes which may be associated with outflows (wind or jet): NLSy1 and CLAGN.

NLSy1 galaxies constitute a class of AGN characterised by the FWHM of the H$\beta$ broad emission line $<$ 2000\,km\,s$^{-1}$ and the flux ratio [O~{\sc iii}]/H$\beta < 3$ \citep{osterbrock85}. NLSy1 are generally believed to have lower BH masses (10$^6$–10$^8$\,$M_{\odot}$) and higher Eddington ratios (e.g. \citealt{smita00,komossa18,rakshit17}). The optical lines, the soft and steep X-ray spectrum ($\Gamma_{X} >2$), the strong Fe~{\sc ii} emission lines and the high amplitude, rapid X-ray variability of NLSy1 reconcile with the observed properties of {\avd}. However, the evolution of this transient is much more complex, such as the disappearance of Fe~{\sc ii} lines and the presence of the Bowen lines \citep{Malyali2021}, as well as the two optical flares. These properties do not strictly conform to the definition of the NLSy1 class.

CLAGN are another type of transients that occur at the centre of active galaxies (typically from a type-1 to type-2 or vice versa) where an evolution of emission line intensity and broadening, as well as high-energy luminosity, are interpreted as caused by a change of the accretion state or Compton-thickness of the AGN (e.g. \citealt{Goodrich1995,Guainazzi2002,LaMassa2015,Ricci2020}). 
Broad Balmer emission lines have been observed by \cite{Malyali2021} in both of the optical flares of {\avd}, which show no significant evolution in the line profiles. The bolometric luminosity and the shape of the X-ray spectrum, however, show significant changes during the second flare (see Fig.~\ref{fig:avd_sed}). The luminosity decreases for nearly two orders of magnitude in 264\,d as the flare evolves from the peak to the late time and the X-ray spectrum hardens. 
Although a typical AGN spectrum is usually hard
with emission up to tens of keV, there are exceptional transient AGN that emit only soft X-rays, e.g. ZTF18aajupnt/
AT2018dyk \citep{frederick19} and ASASSN--18el/AT2018zf/1ES~1927+654 for a certain period when the corona is argued to be disrupted \citep{Ricci2020}. A potential model to explain the absence of hard X-rays in AGN is that when the gas density is high, $> 10^{-14}\, \rm g\,cm^{-3}$, the bremsstrahlung losses exceed the Compton losses for the hot electrons and so they cool very efficiently, suppressing the emission of hard coronal X-rays \citep{Proga2005}.  

However, the evidence, such as the steady IR emission and the quenching of the flaring activity after hundred of days, argue against the existence of a luminous AGN at the location of \avd, at least for the period when {\avd} appeared. Although follow-up monitoring of this source would help to further test this scenario.

A TDE scenario is another possible identification of {\avd} with typical soft X-ray spectra and a peak optical--X-ray luminosity of a few $10^{43}\rm\,erg\,s^{-1}$. 
The radio luminosities of {\avd} -- derived from the different archival and proprietary data -- range between $\sim(0.4$--$7.3) \times 10^{37}$\,erg s$^{-1}$, which are consistent with other radio-detected (jetted) TDEs (10$^{36}$--10$^{42}$\,erg s$^{-1}$, \citealt{Alexander2020}). The delayed radio emission with respect to the optical/X-ray flares is possibly the result of the transition in accretion state similar to other TDE candidates \citep{horesh21a,horesh21b,cendes22}. The IR luminosity of {\avd} is higher than most of TDEs, as well as the covering factor of the dusts \citep{Velzen2016,Jiang2021}. The relatively high luminosity could be due to the contribution of the luminous soft X-rays/EUV. The covering factor is in fact still consistent with the prediction of a 1D radiative transfer model about the dust echoes in TDEs \citep{Lu2016}. Moreover, this value also suggests the circumnuclear environment of \avd is dusty and perhaps gaseous as well.

The most challenging part of the TDE scheme for {\avd} would be its second optical flare. Such two consecutive and non-identical optical flares, lasting for over 1000\,d, have never been observed in the TDE literature. The optical/UV emission in TDEs can originate from different processes when being observed at different phases. At early times, they can be produced by the self-interaction of debris streams \citep{Shiokawa2015,Gezari2017}; at late times, they can be emitted from the outer region of an accretion disc \citep{Velzen2019,Mummery2021}. The X-rays reprocessing process can occur in different phases, when the X-ray emission is being reprocessed either by an optically-thick envelope formed by unbound debris or by outflows \citep{Loeb1997,Roth2016,Gezari2017,Wevers2019}.
In \avd, the first flare could be due to circularization paradigm but the second one should be produced by a different process as it was observed when the first one nearly quenched. The reprocessing can also be ruled out by the facts: i) while the X-rays first declined in MJD~59110--59200, the optical/UV showed a plateau rather than a decreasing trend in the photometries (see more explanation in Sect.~\ref{sec:lag}); ii)  the flare declined earlier in the optical/UV than in the X-rays.

The late-time (at least 5\,yr apart from its peak) UV emission in several TDEs has been reported by \cite{Velzen2019}, who suggest a viscously spreading, unobscured accretion disc model to be responsible for the detection. \cite{Mummery2021} also proposed a unified model with an evolving relativistic thin disc to explain UV and X-ray behaviours in TDEs. However, such late-time UV emission is relatively faint compared to the peak of the flare, only contributing 0.2--5.6\% of the total bolometric energy of a TDE \citep{Velzen2019}. The second flare of {\avd} is apparently much more luminous than this. Though the second plateau present in $\sim$MJD~59483--59580 of the UV lightcurve (see Figs.~\ref{fig:lc} and \ref{figA:UVOT_ph}) is similar to the late-time UV plateau studied by \cite{Velzen2019}.

\cite{Chen2022} proposed a two-phase TDE scenario, in which the first flare of \avd is due to the stream self-collision (i.e. circularization) and the second flare due to a delayed accretion. 
For the origin of optical/UV emission of the second flare, they hinted the reprocessing by an outflow or being directly emitted from the outer region of the late-forming disk, though no details were provided.
Compared to their work, we propose an accretion disc scenario with three distinct accretion rates to further elaborate the origins of the overall multiwavelength emission during the second flaring episode as follows: a compact, hot inner disc region with the highest accretion rate, emitting mainly in soft X-rays; a middle region possibly obscured by outflows; and a truncated outer disc region with a mild accretion rate, dominated by UV and relatively weak optical emissions. Moreover, the optical emission lines were produced by UV reprocessing from the gas locating at sub-pc away from the disc.
A drawback of this picture is how a late-formed accretion disc emits at a super-Eddington rate $\sim$600--800\,d (see Sect.~\ref{sec:xray}) after a TDE.


More exotic scenarios, such as a stellar binary TDE candidate and a TDE disrupted by a SMBH binary, were discussed for {\avd} in \cite{Malyali2021}. 
By statistically studying 70 X-ray selected TDE candidates, \cite{Auchettl2017} found that the column densities of binary TDE candidates are highly enhanced with respect to their Galactic absorption along the line of sight and persist without decreasing beyond 1.5\,yr. 
This result is in good agreement with the TDE scenario that the X-ray is being reprocessed into longer wavelengths at early times of the event.
However, different from the above candidates, the column density of {\avd} is relatively low, consistent with the Galactic absorption in the direction of the transient. This implies that the X-ray flare is unexpected to be observed soon after a TDE.
Even if considering an extreme case in which we viewed nearly through the optically thin funnel region \citep{Dai2018,Thomsen2022}, where the inner disc is mostly exposed to us, there should have been strong X-ray emission from the first flare as well. However, evidence such as lack of Bowen lines in the first flare and the further enhanced IR emission associated with the second flare do not support this scenario. 
In summary, it seems very unlikely that a second TDE caused the second optical flare of {\avd}.

In addition, the combination of multi-band brightening, strong broad and narrow emission lines, and slow decay of luminosities constitutes a new class of transients occurring in the galaxy centre, with the best-studied example being AT~2017bgt \citep{Trakhtenbrot2019}. A `rejuvenated' SMBH which experiences a sudden enhancement or re-ignition of their accretion would produce an intense UV/optical emission. More specifically, AT~2017bgt has shown prolonged enhanced multi-band emission from optical to X-ray with little variability for over 400\,d, a hard X-ray spectrum with a photon index of $\sim1.9$ and a double-peak emission line around 4680\,$\AA$ \citep{Trakhtenbrot2019}. Even though \avd has also shown optical enhancement for over 900\,d accompanied with strong X-ray emission, it has shown much higher and faster variability -- multiple rises and declines -- in its photometry, combined with its ultra-soft X-ray spectrum, both of which make it very different from AT~2017bgt.

\section{Conclusions}
{\avd} has been monitored by ZTF since 2019 and its long-term photometry shows two consecutive flares but with different shapes. X-ray observations performed during the second flare reveal luminous and ultra-soft X-ray emission. We report for the first time the radio (VLA and VLBA) detections of this transient near the peak of the second flare and when the flare was nearly quenched, respectively. 

We propose that the UV and X-rays are produced in the outer and inner regions of the accretion disc, respectively, while the optical and IR emission detected by ZTF (likely line-dominated) and \wise are likely produced by the reprocessing occurred in the gas and dust residing in the circumgalactic medium of the host galaxy.
The $\sim$90--130-d delay between the UV and X-ray decay requires the disc to be geometrically thick, optically thin with a super-Eddington accretion rate, which then moves to a sub-Eddington regime at later times  (see  schematics of the multiwavelength emission in Fig.~\ref{fig:schemetic}).  The radio properties change between the two disc regimes, possibly suggesting an evolving outflow related to the BH accretion (e.g. winds/jets).
The observed Eddington-scaled peak luminosity of $\sim$0.3\,$L_{\rm Edd}$ could be underestimated (e.g. if a portion of the photons are emitted in the unobserved EUV band, if some (undetected) outflows carry materials away, and/or if the mass of the SMBH is overestimated).

We highlight three crucial epochs of the temporal evolution of this transient which we interpret as follows:
 \begin{enumerate}[label={ \arabic*}., leftmargin=0.5em]
 \item First flare: the ignition of the BH activity;
 \item Second flare: formation of a slim disc flaring in from optical to X-ray bands with super-Eddington luminosity, associated with a steep optically-thin radio spectrum.
 \item Post-flare: the accretion disc evolves to a sub-Eddington low-luminosity state,  associated with a compact optically-thick radio emission.
\end{enumerate}

{\it Which type of transient could simultaneously explain these three stages?} A TDE is the most plausible scenario to account for the properties of the first flare. The successive convoluted evolution of {\avd} with a second flare with a long, luminous plateau and the following luminosity decrease do not simply fit in a single class of transients. The possible accretion disc formation
at the second flare and the disc transition to a low state with the consequent ejection of a compact outflow could reconcile with i) a specific type of jetted TDEs with particularly evolving disc properties
\citep{ricarte16,Gezari2017,coughlin19} or ii) a class of AGN which change their disc-jet coupling along their duty cycle (e.g., \citealt{czerny15,davis20,ontiveros21,moravec22}). However, both interpretations fail in explaining some of the multi-band properties of \avd, e.g. the long double-peaked optical lightcurve and the late-time bright X-ray emission in case of a TDE, or the soft X-ray spectrum in case of an AGN. Further theoretical work and long multiwavelength monitoring of {\avd} are needed to verify these interpretations.

\section*{Acknowledgements} 
We thank the anonymous referee for careful reading of the paper and the useful comments.
The authors thank G. Bruni for his support on the radio calibration and data re-processing. The authors thank Pu Du, Shuang-Nan Zhang, Jianmin Wang and Lian Tao for the discussion.
YW acknowledges support from the Royal Society Newton
Fund. RDB acknowledges the support from PRIN INAF 1.05.01.88.06 `Towards the SKA and CTA era: discovery, localisation, and physics of transient sources'. 
XLY thanks the support by Shanghai Sailing Program (21YF1455300), China Postdoctoral Science Foundation (2021M693267) and the National Science Foundation of China (12103076). HY acknowledges support from the Youth Innovation Promotion Association of the CAS (No. 2019060). CJ acknowledges the National Natural Science Foundation of China through grant 11873054, and the support by the Strategic Pioneer Program on Space Science, Chinese Academy of Sciences through grant XDA15052100.

\section*{DATA AVAILABILITY}
The scientific results reported in this paper are based on proprietary VLA (20A-514) and VLBA (BW142A, BW142B) data and archival observations made by VLASS, \wise, ZTF, \sw and \Ni. 

\bibliographystyle{aasjournal}
\bibliography{cite}

\appendix
\counterwithin{figure}{section}

\section{UVOT host subtraction}\label{sec:app}

We compile the host galaxy (2MASX J08233674+0423027) SED using archival observations in the UV through mid-IR bands, we choose magnitudes derived from extended apertures, which for most catalogs are defined at the petrosian radius of the galaxies ($\sim 10\arcsec$ for this host). In the mid-IR we use \wise \citep{Cutri_13} W1 and W2 magnitudes In the near-IR we use UKIDSS \citep{Lawrence_2007} K, H, J and Y bands. We use SDSS DR12 \citep{Alam2015} magnitudes in u, g, r, i, and z optical bands. Finally, for the UV we perform aperture photometry in the GALEX \citep{Bianchi_11} NUV and FUV images with the \textsc{gPhoton} package \citep{Million_16} using a 10\arcsec aperture.

To estimate the host galaxy properties and its brightness in the UVOT bands, we model the SED using the flexible stellar population synthesis \citep[FSPS, ][]{Conroy_09} module. We use the \texttt{Prospector} \citep[][]{Johnson_21} software to run a Markov Chain Monte Carlo (MCMC) sampler \citep[][]{Foreman-Mackey_13}. We assume an exponentially decaying star formation history (SFH), and a flat prior on the five free model parameters: stellar mass ($M_{*}$), stellar metallicity ($Z$), $E(B-V)$ extinction index \citep[assuming the extinction law from][]{Calzetti_2000}, the stellar population age ($t_\mathrm{age}$) and the e-folding time of the exponential decay of the SFH ($\tau_{\rm{SFH}}$).

\begin{figure*} 
\centering  
\includegraphics[width=0.7\linewidth]{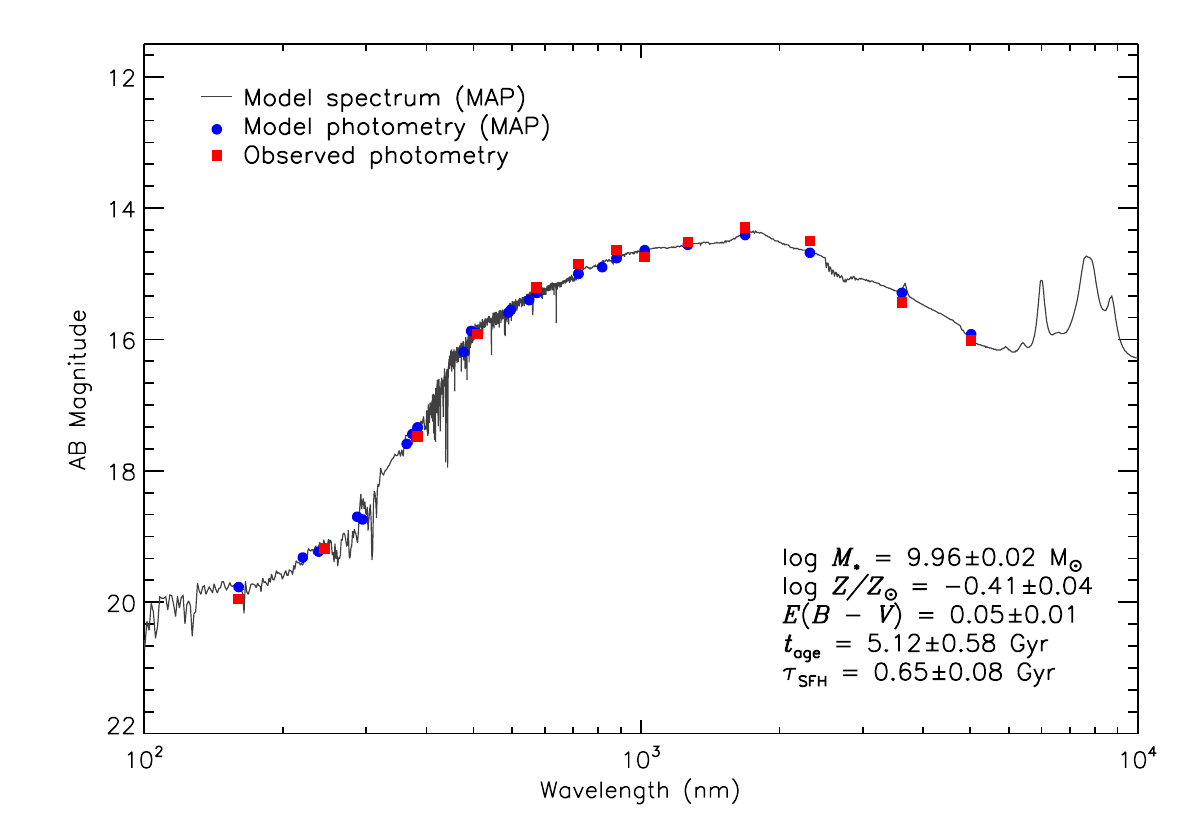}
\includegraphics[width=0.7\linewidth]{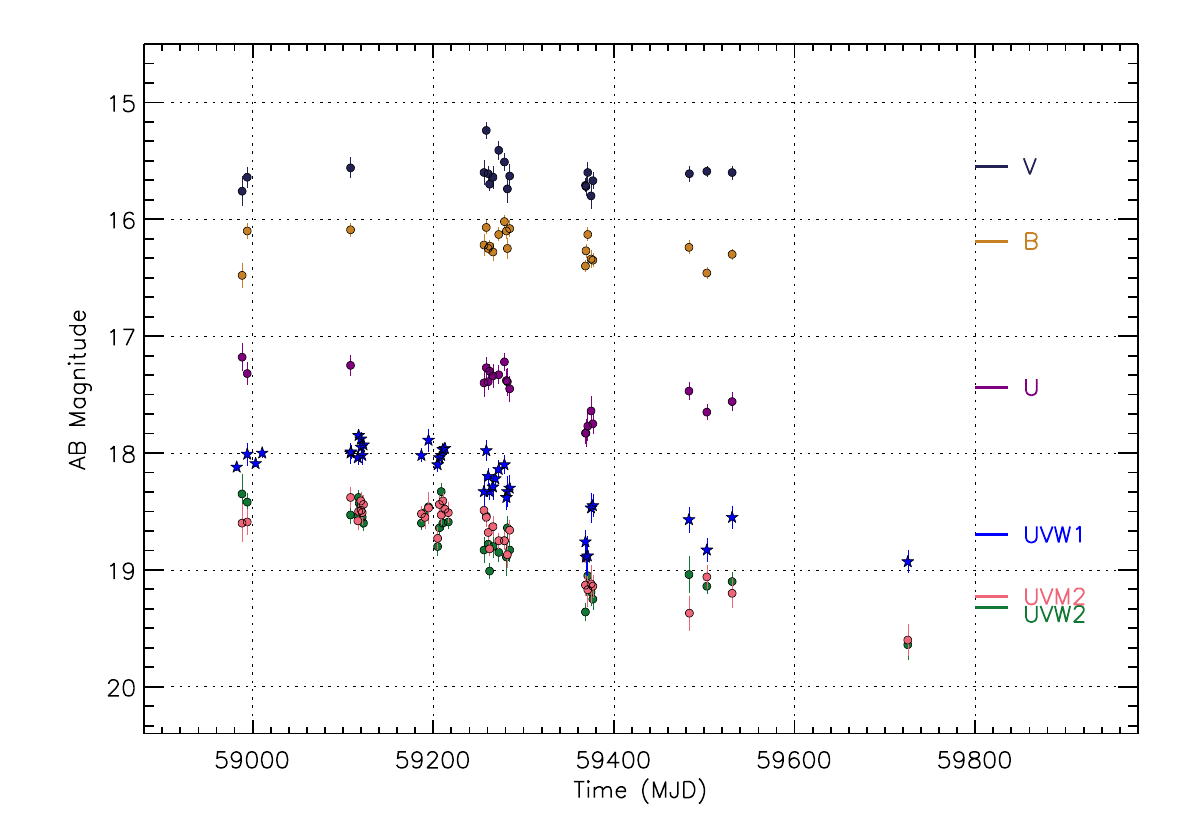}
\vspace{-0.3cm}
\caption{\textbf{Upper:} Host galaxy spectral energy distribution and best-fit template used to synthesise host galaxy magnitudes in the UVOT bands. \textbf{Lower:} UVOT photometries including the contribution from the host galaxy. The lines indicate the magnitude level of the host galaxy for each UVOT filter. The UVW1 data marked with stars are used to represent the UV evolution in Fig.~\ref{fig:lc}.}
\label{figA:UVOT_ph}
\end{figure*}

Using the median and 1-$\sigma$ confidence intervals of the posteriors we derive the host properties, which are shown in the upper panel of Fig.~\ref{figA:UVOT_ph}, alongside the observed SED and modeled SED/spectrum.
We estimate the host galaxy fluxes in the UVOT bands from the posterior distribution of the population synthesis models. The host contribution (see lines in the upper panel of Fig.~\ref{figA:UVOT_ph}) was then subtracted from the observed photometry, which was also corrected for foreground Galactic extinction. The uncertainty on the host galaxy model was propagated into our measurement of the host-subtracted photometry shown in the lower panel of Fig.~\ref{figA:UVOT_ph}.

\section{Best-fitting parameters of the SED in different epochs}\label{sec:best_fitting_para}
We have fitted the SED from IR to soft X-ray with three individual blackbody components and show the best-fitting parameters in Table~\ref{fit_tab}. 
The uncertainties of each parameters were estimated using Monte Carlo simulations and are quoted at 1$\sigma$ confidence level.
\input{fitting_para}

\section{Non-thermal IR emission}\label{sec:non_thermal_IR}
If we assume that the IR emission is non-thermal, then it would have to be synchrotron radiation produced by relativistic electrons that are either more energetic than those producing the radio emission, or that are in a stronger ambient magnetic field. These electrons should cool down quicker than the ones emitting in the radio, which could potentially be related with the steepening of the IR photon index (see Fig.~\ref{fig:WISE}). The transition from a flat spectral index to a negative one could also suggest that the emitter is expanding, thus becoming more transparent to its own emission. 

To properly assess whether this is feasible, we model the IR emitter as a magnetised non-relativistic blob. The conditions that this model needs to take into account are: i) the IR flare duration time in the observer frame ($t_\mathrm{IR} \sim 300$\,d); ii) the frequency of the emitted photons ($\nu_\mathrm{IR} = 8.8\times10^{13}$\,Hz); iii) the peak brightness of the flare ($F_\mathrm{IR} \sim 6$\,mJy); and iv) that a bright ($\sim$ few mJy) flare is not simultaneously seen in the radio band. 
Both the synchrotron cooling time of a particle, $t_\mathrm{syn} \propto B^2 \, E_\mathrm{e}^{-1}$, and the characteristic frequency of the emitted photons, $\nu_\mathrm{syn} \propto B \, E_\mathrm{e}^2$, depend on the energy of the electron, $E_\mathrm{e}$, and the magnetic field intensity in the emitter, $B$ \citep{Blumenthal1970}. 
We assume that the duration of the IR flare is given by the electron synchrotron cooling time, and that the frequency of the synchrotron photons matches the frequency of the observed IR photons. Thus, conditions i) and ii) become 
$t_\mathrm{syn}(E,B) = t_\mathrm{IR}$ and $\nu_\mathrm{IR} = \nu_\mathrm{syn}$, from where we can derive the magnetic field intensity in the jet and the energy of the emitting electrons ($B \approx 23$\,mG and $E_\mathrm{e} \approx 300$\,MeV, respectively). This calculation implicitly assumes that $B$ does not vary significantly during the flare.
In addition, one can assume that the observed flux is due to optically thin synchrotron emission from a relativistic electron population of the form $N(E) dE = N_0 E^{-p} dE$, with $p\sim2.2$. We further assume that the minimum and maximum energies of the electrons are 1 MeV and 1 TeV, respectively, and normalise the electron distribution by adopting an energy condition, $U_\mathrm{e} = \eta_B U_B$, which ties the energy densities in relativistic electrons and the magnetic field via the factor $\eta_B$ ($\eta_B=1$ corresponds to energy equipartition, usually assumed as a reference case). With that, we can obtain $N_0$ from $U_\mathrm{e} = \int_{E_\mathrm{min}}^{E_\mathrm{max}} E N(E) dE$, and use it to calculate the optically-thin synchrotron SED at any given frequency \citep{Blumenthal1970}. Considering that the emitter is homogeneous and spherical, from condition iii) we can estimate its size, $R_0$. We do this for values of $\eta_B$ in the range 0.01--100. We obtain $R_0 \approx 7\pm 6 \times10^{18}$\,cm. However, this large size of the emitter contradicts condition iv), as it leads to a synchrotron-self absorption opacity $\tau_\mathrm{SSA} \ll 1$ even at radio frequencies. The small opacity means that the putative IR synchrotron spectrum should extend to radio frequencies with the same spectral index, yielding a flux $\gg 10$\,mJy. This is a strong argument against this synchrotron interpretation.

\end{document}

%% file: table_radio.tex
\begin{table*}
\caption{Radio observations of AT2019~avd}
    \centering
    \begin{tabular}{ccc|cccccccc}
    \hline
    Instrument & $\nu$  & $\Delta\nu$ & MJD & $\alpha$(J2000) & $\delta$(J2000)  & $\theta_{\rm M}\times\theta_{\rm m}$  & PA & rms  &  F$_{\rm peak}$ & F$_{\rm tot}$ \\
    \hline
    FIRST & 1.4 & 0.042  &  51955.0    &                 &         &   &  & 146  &  $<$438 &    \\      
    VLASS & 3.0 & 2  & 58075.0       &               &               &   &   &  130  & $<$390  &  \\ 
    VLASS & 3.0 & 2  & 59070.8       &            &             &   &  & 169  &  $<$506 &    \\
    \hline
    VLA  & $10.0$ & 4   & 58994.9    & 08:23:36.764 &  04:23:02.481  & 1.10$\arcsec\times$0.57$\arcsec$  &   6.7  &  8.5  &  279$\pm$10 &  322$\pm$14  \\
    VLA  & 9.0$^\natural$  &2  & -   & 08:23:36.764 &  04:23:02.434 &  1.45$\arcsec\times$0.87$\arcsec$  & 11  & 10  &  300$\pm$12 &  373$\pm$14 \\
    VLA  & 11.1$^\natural$ &2  & -   & 08:23:36.762  &  04:23:02.427   & 1.13$\arcsec\times$0.27$\arcsec$ &  8.5 &  10  &   247$\pm$11 & 266$\pm$13 \\
    \hline
    VLBA & 1.55 & 0.256 &   59387.9     &   08:23:36.7666 & 04:23:02.5032   &  11.7$\times$4.8 mas$^2$ &  $-0.91$ & 25   & 173$\pm$26 & 170$\pm$38 \\
    VLBA & 4.98 & 0.256 &   59448.8     &  08:23:36.7666  & 04:23:02.5037   &  3.7$\times$1.6 mas$^2$  &  1.44    & 16   & 566$\pm$35  & 625$\pm$38 \\
    \hline
    \end{tabular}
    \begin{flushleft}
  \textbf{Column description:} (1) instrument, (2) central frequency (GHz), (3) bandwidth (GHz); (4) MJD observation date; (5-6) coordinates of the radio centre position (J2000.0); (7) deconvolved FWHM dimensions (major $\times$ minor axes, $\theta_{\rm M}\times\theta_{\rm m}$) of the fitted component, determined from an elliptical Gaussian fit; (8) PA of the deconvolved components (degree) ; (9) rms of the radio map close to the specific component ($\mu$Jy\,beam$^{-1}$); (10) peak brightness in $\mu$Jy\,beam$^{-1}$, F$_{\rm peak}$; (10) integrated flux density, $F_{\rm tot}$, in $\mu$Jy derived from the \textsc{casa} gaussian fitting of the radio core. $^\natural$ indicates that we divided the 10\,GHz VLA observation (4-GHz wide) into two sub-bands (2-GHz wide).
    \end{flushleft}
    \label{radiotab}
\end{table*}

%% file: fitting_para.tex
\begin{table*}
\caption{Best-fitting parameters of the SED in the three epochs}
\renewcommand{\arraystretch}{1.2}
    \begin{tabular}{lcccccccc}
    \hline
    epoch & MJD & band  &  log$_{10}(kT_{\rm bb})/\Gamma$ & log$_{10}(R_{\rm bb})$ & log$_{10}(F_{\rm bb})$/log$_{10}(F_{\rm pl})$& stat/$\nu$\\
    \hline
& 58994.9 & IR   &  3.05$\pm0.02$   & 17.07$\pm0.02$ & $-11.12\pm0.01$ & -  \\
Pre-peak  & 58994.0 & OUV  & 4.27$\pm0.02$  & 14.32$\pm0.03$ & $-11.73\pm0.06$ & 13.9/3 \\
 & 58994.0 & X-ray   & 6.15$\pm0.02$    & 10.57$\pm0.04$ & $-11.85\pm0.07$ & 17.95/13\\
    \hline
& 59151.9 & IR   & 3.00$\pm0.02$   & 17.14$\pm0.03$ & $-11.18\pm0.01$  &-  \\
Around-peak& 59108.4 & OUV  & 4.17$\pm0.02$ & 14.54$\pm0.04$ & $-11.70\pm0.04$ & 10.0/4 \\
 & 59108.4 & X-ray   &  6.13$\pm0.01$    & 11.10$\pm0.02$ &$-10.89\pm0.02$  &85.4/77  \\
    \hline
 & 59519.1 & IR & 2.96$\pm0.02$ & 17.15$\pm0.04$ & $-11.33\pm0.01$  & -  \\
 Post-flare & 59530.8 & OUV & 3.93$\pm0.25$ & 14.83$_{-0.64}^{+1.26}$ & $-12.10_{-0.39}^{+1.32}$& 3.25/1 \\
& 59483-59725 & X-ray   & $2.09_{-0.54}^{+0.62}$ & - &  $-13.35\pm0.15$  &1.8/7 \\
    \hline
    \end{tabular}
    \begin{flushleft}
  \textbf{Column description:} (1) epoch; (2) MJD observation date; (3) observation band; (4) the blackbody temperature in units of K/the photon index; (5) the inner radius in units of cm; (6) the (un)absorbed flux in units of $\rm erg\,cm^{2}\,s^{-1}$; (7) the goodness of fit. The Cash and $\chi^2$ statistics are applied to assess the X-ray and the optical-to-UV SED modelling, respectively. 
    \end{flushleft}
    \label{fit_tab}
\end{table*}